\documentclass[useAMS,usenatbib,usegraphicx]{mn2e}
\usepackage{bm}
\usepackage{txfonts}

\title[Dust processing in elliptical galaxies]
{Dust processing in elliptical galaxies}
\author[Hirashita et al.]{Hiroyuki Hirashita,$^1$\thanks{E-mail:
    hirashita@asiaa.sinica.edu.tw}
Takaya Nozawa,$^2$ Alexa Villaume$^3$
and Sundar Srinivasan$^1$\\
$^1$Institute of Astronomy and Astrophysics, Academia Sinica,
PO Box 23-141, Taipei 10617, Taiwan\\
$^2$National Astronomical Observatory of Japan, Mitaka, Tokyo 181-8588, Japan\\
$^3$Department of Astronomy and Astrophysics, University of California, Santa Cruz, CA 95064, USA
}
\date{2015 September 14}

\pagerange{\pageref{firstpage}--\pageref{lastpage}} \pubyear{2015}

\begin{document}
\label{firstpage}
\maketitle

\begin{abstract}
We reconsider the origin and processing of dust in
elliptical galaxies. We theoretically formulate
the evolution of grain size distribution, taking into account
dust supply from asymptotic giant branch (AGB) stars
and dust destruction by sputtering in the hot interstellar
medium (ISM), whose temperature evolution is treated
by including two cooling paths: gas emission and dust
emission (i.e.\ gas cooling and dust cooling).
With our new full treatment of grain size distribution,
we confirm that dust destruction by sputtering
is too efficient to explain the
observed dust abundance even if AGB stars continue to
supply dust grains,
and that, except for the case where the
initial dust-to-gas ratio in the hot gas is as high as $\sim 0.01$,
dust cooling is negligible compared with gas cooling.
However, we show that,
contrary to previous
expectations, cooling does not help to protect the
dust; rather, the sputtering efficiency is raised by
the gas compression as a result of cooling. We additionally
consider grain growth after the gas cools down.
Dust growth by the accretion of gas-phase
metals in the cold medium increase the dust-to-gas ratio up to
$\sim 10^{-3}$ if this
process lasts $\ga 10/(n_\mathrm{H}/10^3~\mathrm{cm}^{-3})$~Myr,
where $n_\mathrm{H}$ is the number density of
hydrogen nuclei.
We show that the accretion of gas-phase metals is a viable
mechanism of increasing
the dust abundance in elliptical galaxies to a level consistent
with observations, and that
the steepness of observed extinction curves is
better explained with grain growth by accretion.
\end{abstract}

\begin{keywords}
dust, extinction ---
galaxies: elliptical and lenticular, cD ---
galaxies: evolution ---
galaxies: ISM ---
methods: analytical
\end{keywords}

\section{Introduction}

In the nearby Universe, elliptical galaxies are known
to have little ongoing star formation activity. However,
elliptical galaxies are not completely devoid of gas:
they are also known to have
hot X-ray emitting halo gas \citep[e.g.][]{osullivan01} and cold gas
\citep[e.g.][]{wiklind95}. Moreover,
dust is detected for a large ($>$50 per cent) fraction of
elliptical galaxies by optical extinction
\citep[e.g.][]{goudfrooij94a,vandokkum95,ferrari99,tran01} or
far-infrared (FIR) emission \citep[e.g.][]{knapp89}.
Dust mass is estimated to be $10^4$--$10^5$~M$_{\sun}$ from the
reddening in the optical \citep{goudfrooij94a}. The dust
mass estimated from FIR dust emission tends to be even
larger ($\sim 10^5$--$10^7$ M$_{\sun}$; e.g.\ \citealt{leeuw04,temi04}).
The existence of dust is also important in determining the
chemical properties of the cold gas through dust-surface
reactions and freeze-out of molecules onto the dust
\citep*{fabian94,voit95}. In spite of such an important role of dust,
the origin of dust in elliptical galaxies is still
being debated, and there is no clear theoretical explanation
for the total amount of dust there.

Because the stellar population is
dominated by old stars whose ages are comparable to the
cosmic age, the dust is predominantly supplied by asymptotic giant
branch (AGB) stars in elliptical galaxies. However,
the supplied dust is quickly
destroyed by sputtering in the X-ray-emitting
hot gas. This destruction is so efficient that
the observed dust mass cannot be explained by the
dust abundance achieved by the balance between the supply
from AGB stars
and the destruction \citep[e.g.][]{patil07}.
Therefore, the existence of such an `excessive' amount of dust
has long been a mystery.

Some authors argue that the dust existing in elliptical galaxies
is possibly injected from outside via the merging or accretion of
external galaxies \citep{forbes91,temi04,fujita13}.
The lack of correlation between dust FIR luminosity and
stellar luminosity is also taken as evidence of
external origin of dust \citep*{temi07}; however, this argument
implicitly assumes that the existing dust is tightly related
to the stellar dust production activity.
If dust is processed by mechanisms not
related to stars, it may be natural that we do not find a correlation
between dust emission and stellar luminosity. In particular,
dust may grow through the accretion of gas-phase metals
in the cold gas, and this dust growth is suggested to be the
most efficient mechanism of dust mass increase in nearby
galaxies
\citep*[e.g.][]{dwek98,hirashita99,zhukovska08,inoue11,mattsson12,asano13a,debennassuti14,remy14}.
Indeed, \citet{fabian94} and \citet{voit95} pointed out a possibility
of dust reformation in a cooling flow.
More recently, \citet{martini13} also suggested the importance
of accretion in maintaining the dust abundance to a level
consistent with observations by analogy with the Milky Way.
Since the efficiency of dust growth is not necessarily
related to the stellar properties
of elliptical galaxies, the lack of correlation between
dust and stellar emissions does not necessarily mean
the external origin of dust.


The interstellar dust is not only passively processed in the ISM,
but could also actively regulate the physical state of the ISM,
since dust is able to radiate away the energy obtained by
collisions with gas particles. This is one of the most important
cooling paths of the hot ISM, and referred to as dust cooling.
\citet{mathews03} argued that dust cooling could occur faster
than the dynamical time in the galactic potential and thus
possibly enable the dust to
survive against sputtering. However, they did not treat the dust
supply from AGB stars consistently, and they assumed a
relatively high initial
dust-to-gas ratio ($\sim$0.01, which is comparable to the dust
rich ISM seen in the Milky Way) in the hot gas. Since there has been some
advance in the theoretical understanding of dust production in
AGB stars \citep{ferrarotti06,ventura12}, it is worth reconsidering
their models using
the new knowledge about the dust yield.
Moreover,
since \citet{mathews03} assumed a single grain size for dust cooling,
it would be interesting to calculate the evolution of grain
size distribution: indeed, the grain size distribution is of
fundamental importance in comparison with observed
extinction curves as we argue below. The rates of sputtering and
dust cooling also depend on the grain size distribution
\citep*{dwek87,nozawa06}.

The evolution of dust is also linked to active galactic nucleus (AGN)
activities in elliptical galaxies.
\citet{temi07} proposed the transportation of dust from dust reservoirs
in the central regions to explain the extended dust emission
in elliptical galaxies \citep[see also][]{mathews13}.
In this context, the origin, production, and survival of dust
are important in clarifying the origin of the dust reservoirs.
As mentioned above, dust may also contribute to cooling.
The cool gas that falls into the
centre may fuel the AGN \citep[e.g.][]{werner14},
which in turn works as a heating
source and reforms the hot gas \citep{ciotti01}.
Recently, \citet{valentini15} pointed out that the
AGN feedback could also enhance cooling locally through compression.

In this paper, we focus on the processing of dust in elliptical
galaxies, both in the hot and cold gas components.
In order to avoid complexity
arising from the AGN feedback, we do not treat the supply of
hot gas through the feedback but start the calculation given
the pre-existing hot gas.
Instead, we do treat cooling of the gas in a consistent
manner with the
dust supply and destruction in the hot gas.
Since dust cooling and dust destruction depend on the
grain size distribution, we fully take the evolution of grain size
distribution into account.
We additionally consider the impact of dust growth in
the cooled gas not only on the dust amount, but also on
the grain size distribution.
Through the modeling of the above processes,
we will be able to clarify the evolution and survival of dust
in a cooling cycle of the ISM in elliptical galaxies.

We emphasize that the treatment of grain size distribution
is one of the most important features in our modeling.
Indeed, there is an observational clue to the
grain size distribution, which enables us to test our results.
The wavelength dependence of extinction, the so-called
extinction curve, is investigated for a statistical number
($\ga$10) of elliptical
galaxies: \citet{goudfrooij94b} showed that the ratio of
total to selective extinction $R_V$, which is a measure of
the flatness of optical extinction curve,
tends to be smaller ($R_V=2.1$--3.3) than the mean
value of the Milky Way extinction
curves (3.1;
e.g.\ \citealt{pei92,weingartner01,draine03,fitzpatrick07,nozawa13}).
This implies that
the typical grain size is smaller
in the elliptical galaxies than in the Milky Way.
\citet{patil07} obtained a similar range ($R_V=2.03$--3.46)
for 11 elliptical galaxies.
These extinction curves can be used to
examine the evolution of grain size distributions through
various processes in the ISM.

The paper is organized as follows: we explain our models
for the basic processes of dust evolution in the hot
ISM in Section \ref{sec:model}. We show the
results in Section \ref{sec:result}. We additionally consider
dust processing in the cold ISM, focusing on dust growth,
in Section \ref{sec:cool}. In Section \ref{sec:obs}, some
predicted features of the models such as dust abundance and
extinction curves are compared with observations.
Finally we conclude in Section \ref{sec:conclusion}.

\section{Processing in the hot ISM}\label{sec:model}

We explain the method of calculating the
evolution of dust grain size distribution in the hot ISM
in an elliptical galaxy. Our models are composed of
dust supply by AGB stars and dust destruction by
sputtering. Cooling of the hot gas by dust thermal
emission and gas line emission is also calculated
in a consistent manner with the evolution of dust
grain size distribution. To avoid the complexity arising
from dynamical evolution determined by
the gravitational field and the AGN heating,
we only consider the evolution of a local `fluid element'
of the ISM without modeling the global
dynamics.
The simplicity of this approach enables us to
focus on dust processing which is determined by the
local environment where the dust resides.

\subsection{Evolution of grain size distribution}

The grain size distribution at time $t$ is defined so that
$n(a,\, t)\,\mathrm{d}a$ is the number density of grains
in the range of grain radii between $a$ and $a + \mathrm{d}a$.
The time $t$ is measured from the onset of dust processing
in the hot gas ($t$ is \textit{not} the age of the galaxy).
As mentioned above, we focus on a local region without
considering the global gas and dust distributions within the
galaxy. We also assume that the gas and dust
are well coupled (i.e.\ once dust is injected into a certain
gas element, that dust stays in the same element without any
displacement; in other words, the Lagrangian trajectories
of gas and dust are the same).
Under these assumptions, the evolution of
grain size distribution in the hot gas is described as
\begin{eqnarray}
\frac{\upartial n(a,\, t)}{\upartial t}+
\frac{\upartial}{\upartial a}[\dot{a}n(a,\, t)]=S(a,\, t)+
n(a,\, t)\frac{\mathrm{d}\ln\rho_\mathrm{gas}}{\mathrm{d}t},
\label{eq:continuity}
\end{eqnarray}
where $\dot{a}=\mathrm{d}a/\mathrm{d}t$ is the
changing rate of grain radius by sputtering
($\dot{a}$ is negative; see Section \ref{subsec:sput}),
$S(a,\, t)$ is the source term due to the dust production by
AGB stars (Section \ref{subsec:star}), and $\rho_\mathrm{gas}$
is the gas mass density.
The last term in equation (\ref{eq:continuity}) is
the increase of the grain number density by
the change of the background gas density (Appendix \ref{app:continuity}):
as formulated in Section \ref{subsec:cooling},
we consider the change of gas density by cooling.
We give the number density of
hydrogen nuclei, $n_\mathrm{H}$, as an input parameter,
and the relation between $n_\mathrm{H}$ and $\rho_\mathrm{gas}$ is
given by
\begin{eqnarray}
\rho_\mathrm{gas}=\mu m_\mathrm{H}n_\mathrm{H},\label{eq:dens}
\end{eqnarray}
where $m_\mathrm{H}$ is the mass of hydrogen atom,
and $\mu$ is the mean weight of the gas per hydrogen
(we adopt $\mu =1.4$).

Since we are often interested in the grain mass,
we introduce the grain size distribution per unit grain
mass by defining $\tilde{n}(m,\, t)\,\mathrm{d}m$ as
the number density of grains with mass between $m$ and
$m+\mathrm{d}m$; i.e.\
$\tilde{n}(m,\, t)\,\mathrm{d}m=n(a,\, t)\,\mathrm{d}a$.
The grain mass is estimated as $m=\frac{4}{3}\upi a^3s$
(i.e.\ grains are assumed to be spherical),
where constant $s$ is the material density of the grain.
For the calculation of grain size distribution, we
adopt $s=3.3$ g cm$^{-3}$ based on a silicate material.
Using the fact that $\dot{a}$ is independent of $a$
(Section \ref{subsec:sput}),
we obtain \citep[see also][]{hirashita12}
\begin{eqnarray}
\frac{\upartial (m\tilde{n})}{\upartial t}+\dot{m}
\frac{\upartial (m\tilde{n})}{\upartial m}=\frac{1}{3}
\dot{m}\tilde{n}+m\tilde{S}(m,\, t)+
m\tilde{n}\frac{\mathrm{d}\ln\rho_\mathrm{gas}}{\mathrm{d}t},
\label{eq:continuity2}
\end{eqnarray}
where
$\dot{m}\equiv\mathrm{d}m/\mathrm{d}t=4\pi a^2s\dot{a}$
and $\tilde{S}(m,\, t)\,\mathrm{d}m=S(a,\, t)\,\mathrm{d}a$.

The dust-to-gas ratio as a function of time, $\mathcal{D}(t)$,
is calculated by
\begin{eqnarray}
\mathcal{D}(t) & = & \frac{1}{\mu m_\mathrm{H}n_\mathrm{H}}
\int_0^\infty\frac{4}{3}\upi a^3sn(a,\, t)\,\mathrm{d}a\nonumber\\
& = & \frac{1}{\mu m_\mathrm{H}n_\mathrm{H}}
\int_0^\infty m\tilde{n}(m,\, t)\,\mathrm{d}m.\label{eq:dg}
\end{eqnarray}

\subsection{Supply from stars}\label{subsec:star}

\subsubsection{Formulation}\label{subsubsec:AGB_formulation}

We consider the stellar population contributing to the
dust supply in the local gas element considered above.
Since we are interested in a system as old as the cosmic age,
the current stellar population contributing to the dust
formation is AGB stars, which originate from stars
with $m<8$ M$_{\sun}$ ($m$ is the mass at the zero-age
main sequence) \citep[e.g.][]{zhukovska08}.
The dust mass density supplied from AGB stars
per unit time, $\dot{\rho}_\mathrm{dust,AGB}$, is estimated as
\begin{eqnarray}
\dot{\rho}_\mathrm{dust,AGB}=\int_{m_{t_\mathrm{age}}}^{m_\mathrm{up}}
m_\mathrm{d}(m)\phi (m)\dot{\rho}_*(t_\mathrm{age}-\tau_m)\,\mathrm{d}m,
\label{eq:supply}
\end{eqnarray}
where $m_{t_\mathrm{age}}$ is the turn-off mass (i.e.\ the mass
of the star whose lifetime is equal to the galaxy age
$t_\mathrm{age}$),
$m_\mathrm{up}$ is the upper cut-off of stellar mass,
$m_\mathrm{d}(m)$ is the total dust mass produced
by an AGB star with mass $m$,
$\phi (m)$ is the stellar initial mass function (IMF),
$\dot{\rho}_*(t_\mathrm{age})$ is the star formation rate density
as a function of galaxy age, and $\tau_m$ is the lifetime of the star
with mass $m$. We take the stellar lifetime from \citet{raiteri96}.
We adopt a Salpeter IMF [$\phi (m)\propto m^{-2.35}$] with
$m_\mathrm{up}=100$~M$_{\sun}$ and $m_\mathrm{low}=0.1$ M$_{\sun}$,
where $m_\mathrm{low}$ is the lower cut-off of stellar mass, and
the normalization of the IMF is determined by
$\int_{m_\mathrm{low}}^{m_\mathrm{up}}m\phi (m)\mathrm{d}m=1$.
The data of dust yield $m_\mathrm{d}(m)$ is described in
Section \ref{subsubsec:yield}.
We assume that dust grains ejected from AGB stars are instantaneously
(on a time-scale shorter than $10^5$ yr; \citealt{mathews99})
injected into the hot gas \citep{temi07}.
We approximate the star formation history
with an instantaneous burst at $t_\mathrm{age}=0$, which is a good approximation
for nearby elliptical galaxies:
\begin{eqnarray}
\dot{\rho}_*(t_\mathrm{age})=\rho_\mathrm{*,tot}\delta (t_\mathrm{age}),
\label{eq:delta}
\end{eqnarray}
where $\rho_\mathrm{*,tot}$ is the total stellar mass density
ever formed and $\delta (t)$ is Dirac's $\delta$ function.
Noting that only stars with mass $m_{t_\mathrm{age}}$ are
contributing to the current dust formation, we obtain
from equations (\ref{eq:supply}) and (\ref{eq:delta})
\begin{eqnarray}
\dot{\rho}_\mathrm{dust,AGB}=m_\mathrm{d}(m_{t_\mathrm{age}})\phi (m_{t_\mathrm{age}})
\rho_\mathrm{*,tot}\left| \frac{\mathrm{d}m_t}{\mathrm{d}t}\right|_{t=t_\mathrm{age}} .
\end{eqnarray}
Taking the mass loss of stars that have ended
their lives into account, the current stellar mass,
$\rho_*$, is estimated as
\begin{eqnarray}
\rho_*=(1-\mathcal{R})\rho_\mathrm{*,tot},
\end{eqnarray}
where $\mathcal{R}$ is the returned fraction. We assume
that $\mathcal{R}=0.25$ \citep{kuo13} (this value is derived
for $m_{t_\mathrm{age}}\simeq 1$ M$_{\sun}$, but
is not sensitive to $t_\mathrm{age}$ as long as
$t_\mathrm{age}>10^9$ yr; \citealt{hirashita11}).
Correspondingly, we adopt $t_\mathrm{age}=9.5$ Gyr,
the lifetime of $m=1$~M$_{\sun}$ star. We fix $t_\mathrm{age}$,
since we are interested in the time evolution at $t\la 10^8$ yr,
which is much shorter than $t_\mathrm{age}$.



Now we formulate the source term in equation (\ref{eq:continuity}).
This term expresses the dust input from AGB stars, and we
hereafter refer to
the dust supplied by AGB stars as \textit{AGB dust}.
We specify the grain size distribution of AGB dust in such
a way that the source term in equation (\ref{eq:continuity})
is written as
\begin{eqnarray}
S(a,\, t)=\dot{\rho}_\mathrm{dust}f_\mathrm{dust}(a),\label{eq:source}
\end{eqnarray}
where $f_\mathrm{dust}(a)$ is the grain size distribution function
whose normalization is determined by
\begin{eqnarray}
\int_{0}^{\infty}\frac{4}{3}\pi a^3sf(a)\,
\mathrm{d}a=1.\label{eq:norm_AGB}
\end{eqnarray}
Or equivalently, it is possible to use equation (\ref{eq:continuity2})
with
$\tilde{S}(m,\, t)=\dot{\rho}_\mathrm{dust}\tilde{f}_\mathrm{dust}(m)$,
where the normalization of mass distribution function
$\tilde{f}_\mathrm{dust}(m)$ is determined by
$\int_0^\infty m\tilde{f}_\mathrm{dust}(m)\,\mathrm{d}m=1$.
For the
grain size distribution, we consider two cases: one is
a power-law form with the same power index as used
in \citet{tsai95},
\begin{eqnarray}
f_\mathrm{dust}(a)=C_\mathrm{M}a^{-3.5}~(0.001~\micron\leq a\leq 0.25~\micron ),
\label{eq:mrn}
\end{eqnarray}
and the other is a lognormal form,
\begin{eqnarray}
f_\mathrm{dust}(a)=\frac{C_\mathrm{l}}{a}
\exp\left\{ -\frac{[\ln (a/a_0)]^2}{2\sigma^2}\right\},
\label{eq:lognorm}
\end{eqnarray}
where $C_\mathrm{M}$ and $C_\mathrm{l}$ are
the normalizing constants determined
by equation (\ref{eq:norm_AGB}), and $a_0$ and $\sigma$
are the central
grain radius and the standard deviation of the lognormal
distribution, respectively.
The first one (equation \ref{eq:mrn})
is based on the one derived
for the Milky Way ISM \citep*{mathis77}, and is referred to as
the MRN grain size distribution. The second one in
equation (\ref{eq:lognorm}) is based on
the argument that AGB stars produce large (0.1--1~$\micron$) grains
\citep[][and references therein]{asano13}. We adopt the same
parameters as in \citet{asano13}: $a_0=0.1~\micron$ and
$\sigma =0.47$, which are based on theoretical calculations
of dust condensation in AGB star winds by \citet{yasuda12}.

\subsubsection{Dust yield of AGB stars}\label{subsubsec:yield}

We take $m_\mathrm{d}(m)$ (the ejected dust mass by
an AGB star for different progenitor mass) from
theoretical calculations by \citet{zhukovska08}, who adopted
\citet{ferrarotti06}'s framework to calculate the
dust condensation in AGB star winds.
Since we are interested in nearby elliptical galaxies,
it is reasonable to consider
the solar metallicity case ($Z=0.02$). The stellar
mass around 1~M$_{\sun}$, which is contributing to the
current dust production, is relevant to this paper.
Their calculations indicate that around
$4\times 10^{-4}$--$10^{-3}$~M$_{\sun}$
of dust is produced by a $m=$ 1--2~M$_{\sun}$ star.
If we adopt $Z=$ 0.015 (0.03) instead, the dust mass
becomes 3 times as small (large) as the case above.
Therefore, we should keep a factor of 3 uncertainty in
mind.
\citet{ventura14} also calculated theoretical dust yield for
AGB stars, focusing on lower metallicities. Their highest
metallicity case ($Z=8\times 10^{-3}$) shows that the
dust mass produced by an AGB star with $m=$ 1--2~M$_{\sun}$
is $2\times 10^{-4}$--$10^{-3}$ M$_{\sun}$, similar to that
adopted above.


\subsection{Sputtering}\label{subsec:sput}

The changing rate of grain radius by sputtering,
$\dot{a}$, in equation (\ref{eq:continuity}) is estimated by
a fitting formula derived by \citet{tsai95}:
\begin{eqnarray}
\dot{a}=-\tilde{h}\left(
\frac{\rho_\mathrm{gas}}{m_\mathrm{H}}\right)\left[1+
\left(\frac{\tilde{T}}{T}\right)^{2.5}\right]^{-1} ,\label{eq:sput}
\end{eqnarray}
where the coefficients of this fitting function,
$\tilde{h}=3.2\times 10^{-18}$ cm$^4$ s$^{-1}$
and $\tilde{T}=2\times 10^6$ K, are determined to give a
good fit to the sputtering rate of both silicate and graphite
in \citet{draine79}.
This formula correctly reproduces the temperature
dependence of sputtering
in that
it is efficient at $T\ga 2\times 10^6$ K and steeply
declines at $T\la 10^6$ K
\citep{nozawa06}.
We also define the sputtering
time-scale, $\tau_\mathrm{sput}\equiv |a/\dot{a}|$,
which is estimated at $T\ga\tilde{T}$
using equation (\ref{eq:dens}) as
\begin{eqnarray}
\tau_\mathrm{sput}\simeq\frac{a}{\tilde{h}\mu n_\mathrm{H}}
\simeq 7.1\times 10^5\left(\frac{a}{1~\micron}\right)
\left(\frac{n_\mathrm{H}}{1~\mathrm{cm}^{-3}}\right)^{-1}~\mathrm{yr}.
\label{eq:tau_sp}
\end{eqnarray}
This is consistent with the values derived by
a detailed treatment of sputtering yield
\citep[][see their fig.\ 2]{nozawa06}.

\subsection{Cooling of the hot ISM}\label{subsec:cooling}

Following \citet{mathews03}, we assume that the gas pressure
in the local region of interest is constant.
The time evolution of the
gas temperature $T$ through isobaric cooling is described
by \citep{mathews03}
\begin{eqnarray}
\frac{\mathrm{d}T}{\mathrm{d}t}=-\frac{2\Lambda_\mathrm{tot}(T)}{5nk},
\end{eqnarray}
where $\Lambda_\mathrm{tot}(T)$ is the cooling rate (energy lost per
volume per time)
including both gas and dust cooling, $n$ is the number density
of gas particles, and $k$ is the Boltzmann constant.
Assuming that H and He are fully ionized, we estimate that
$n=2.3n_\mathrm{H}$.

The cooling rate by gas emission is taken from \citet{sutherland93}.
Specifically, we adopt
$n_\mathrm{t}n_\mathrm{e}\Lambda_\mathrm{N}(T)$
[$\Lambda_\mathrm{N}(T)$ is the cooling function for gas cooling]
for the cooling rate per volume with electron number density
$n_\mathrm{e}=1.2n_\mathrm{H}$ and total number density  of ions
$n_\mathrm{t}=1.1n_\mathrm{H}$.
We stop the calculation when $T$ reaches $10^5$~K
(this time is denoted as $t_5$), below
which the temperature evolution occurs much faster
and the effect of recombination would change $n_\mathrm{e}$
significantly.
Since sputtering does not work at such a low temperature,
the resulting grain abundance and size distribution are
insensitive to the detailed treatment of thermal
evolution at $\la 10^5$ K.

The gas also cools through gas--grain collisions
\citep[e.g.][]{dalgarno72,dwek87,seok15}. The rate of this dust cooling
is estimated as $n_en_\mathrm{H}\Lambda_\mathrm{d}(T)$,
where $\Lambda_\mathrm{d}(T)$ is the cooling function for dust
cooling.
We compute $\Lambda_\mathrm{d}(T)$
based on \citet{dwek87} under the dust abundance
and grain size distribution
calculated in this paper. We again adopt $n_e=1.2n_\mathrm{H}$
and assume silicate as dust species (adopting graphite
instead does not significantly change the cooling rate;
\citealt{dwek87}).

After all, we evaluate the total cooling rate as
$\Lambda_\mathrm{tot}=n_\mathrm{t}n_e\Lambda_\mathrm{N}+
n_en_\mathrm{H}\Lambda_\mathrm{d}$.
We assume a constant pressure, so that the density is
simultaneously varied as
\begin{eqnarray}
\frac{\mathrm{d}\ln\rho_\mathrm{gas}}{\mathrm{d}t}=-
\frac{\mathrm{d}\ln T}{\mathrm{d}t}.
\end{eqnarray}

\subsection{Models and initial conditions}\label{subsec:parameter}

For simplicity, we only focus on a `typical' single region
in an elliptical galaxy by adopting representative gas density
and temperature.
\citet{tsai95} considered three models of elliptical galaxies
which are classified by $B$-band luminosities, $L_B$. The
models are
referred to as Models a, b, and c depending on the
$B$-band luminosity. Although we do not directly use $B$ band
luminosity, the quantities used in the models
($\rho_*$,\,$\rho_\mathrm{gas}$,\,$T$) are related
to it.
In Table \ref{tab:param}, we list the adopted parameters
for each model. The values at the core radius are adopted
for the star and gas densities.
(We convert the central densities given in table 1 of
\citealt{tsai95} to the values at the core radius by
assuming the radial profiles given in that paper.)
The stellar mass density $\rho_*$ is used to estimate the
dust supply rate from AGB stars
(Section \ref{subsubsec:AGB_formulation}), and the
gas density $\rho_\mathrm{gas}$ and gas temperature $T$
are used for the initial conditions.

\begin{table}
\centering
\begin{minipage}{85mm}
\caption{Model parameters adopted from \citet{tsai95}.}
\label{tab:param}
\begin{center}
\begin{tabular}{@{}lcccc} \hline
Model & $L_B$ & $\rho_*$ & $\rho_\mathrm{gas}\,^a$ & $T$\\
 & ($10^{10}$\,L$_{\sun}$) & (g\,cm$^{-3}$) &
 (g\,cm$^{-3}$) & ($10^7$\,K)
\\ \hline
a & 10.6 & $1.19\times 10^{-21}$ & $1.43\times 10^{-25}$
& 0.922\\
b & 3.31 & $1.36\times 10^{-20}$ & $3.01\times 10^{-25}$
& 0.468 \\
c  & 0.976 & $1.83\times 10^{-19}$ & $8.74\times 10^{-25}$
& 0.252\\
\hline
\end{tabular}
\end{center}
\textit{Note.} The densities at the core radius are adopted.\\
$^a$The corresponding hydrogen number densities are
$n_\mathrm{H}=0.0610$, 0.128, and  0.373 cm$^{-3}$
for Models a, b, and c, respectively.
\end{minipage}
\end{table}

We start with no dust ($\mathcal{D}_0=0$, where
$\mathcal{D}_0$ is the initial dust-to-gas ratio),
referred to as \textit{the dust-free initial condition}, or
$\mathcal{D}_0=0.0075$, referred to as \textit{the dusty initial
condition}. The latter value of $\mathcal{D}_0$ is typical of the
solar-metallicity ISM and the same functional form
for the grain size
distribution as equation (\ref{eq:mrn})
[i.e.\ $n(a,\, t=0)\propto a^{-3.5}$] is assumed with the normalization
determined by equation (\ref{eq:dg}).
We expect that the realistic situation lies between
these two extreme conditions for the initial
condition, but that the dust-free
initial condition is more realistic considering that
the dust is rapidly destroyed by sputtering as shown
below. Thus,
we focus on the dust-free initial condition
as a `fiducial' model. As we explained above, we
investigate two cases for the grain size distribution
of AGB dust: MRN and lognormal.

\section{Results}\label{sec:result}

\subsection{Thermal evolution and dust-to-gas ratio}
\label{subsec:tem}

\begin{figure}
\includegraphics[width=0.45\textwidth]{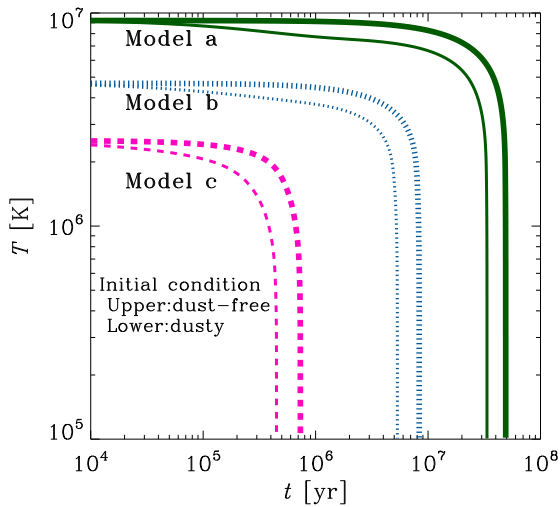}
\caption{Time evolution of gas temperature $T$. Note that the
gas density is always inversely proportional to $T$ during the
entire evolution. The upper thicker (lower thinner)
solid, dotted, and dashed lines present
the results for Models a, b, and c, respectively, with the dust-free
initial condition, $\mathcal{D}_0=0$
(with the dusty initial condition, $\mathcal{D}_0=0.0075$).
The MRN grain size distribution is adopted for
AGB dust. The evolutionary tracks of the cases
without cooling and with the lognormal size distribution of
AGB dust for $\mathcal{D}_0$ both overlap with the
thick lines (thus, they are
not shown in the figure), which means that dust cooling does not
affect the thermal evolution of the hot gas for the dust-free initial
condition.
\label{fig:tem_den}}
\end{figure}

\begin{figure}
\includegraphics[width=0.45\textwidth]{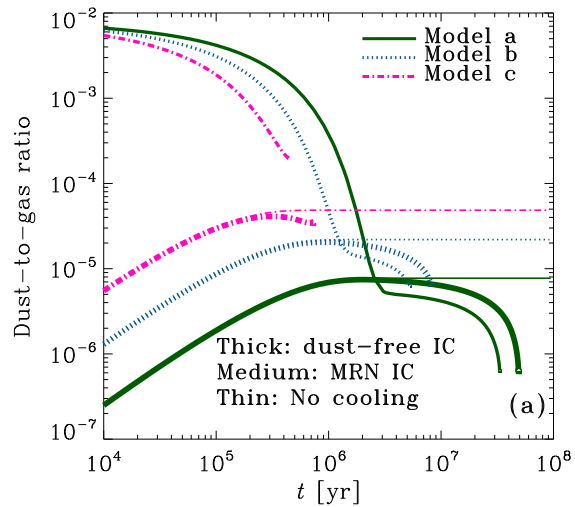}
\includegraphics[width=0.45\textwidth]{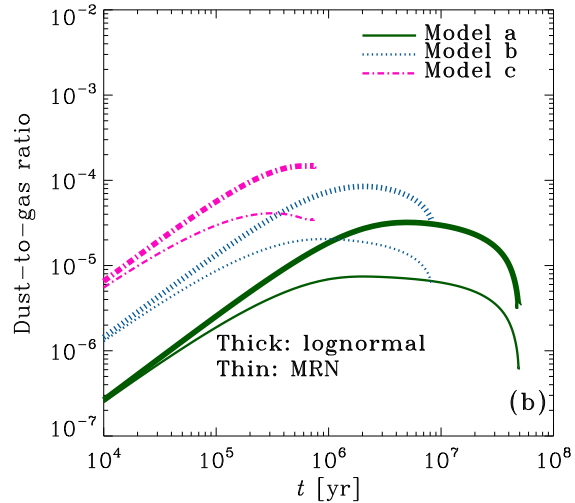}
\caption{Time evolution of dust-to-gas ratio $\mathcal{D}$.
(a) The MRN grain size distribution is adopted for
AGB dust.
The solid, dotted, and dashed lines present
the results for Models a, b, and c, respectively, with the
dust-free and dust-rich initial conditions (ICs) (lines with
enhanced and medium thickness, respectively) with
cooling included, and with the dust-free IC without cooling
(thin lines).  In the cases with cooling,
the calculation is stopped
when the gas temperature reaches $10^5$~K (i.e.\ at time $t_5$).
(b) Comparison between the two different grain size distributions of
AGB dust: lognormal (thick lines) and MRN (thin lines).
The solid, dotted, and dashed lines present
the results for Models a, b, and c, respectively.
\label{fig:dg}}
\end{figure}

The thermal evolution of the hot gas is of fundamental
importance in determining the fate of dust. On the other hand,
the temperature of hot gas could also be affected
by dust cooling. For dust cooling, the dust abundance
is the key factor. Thus, we first show
the time evolutions of
gas temperature ($T$) and dust-to-gas ratio ($\mathcal{D}$)
in Figs.\ \ref{fig:tem_den}
and \ref{fig:dg}, respectively.

\begin{table}
\centering
\begin{minipage}{85mm}
\caption{Time ($t_5$) at which the temperature drops down to $10^5$ K.}
\label{tab:t5}
\begin{center}
\begin{tabular}{@{}lcccc} \hline
IC$^a$ & AGB dust$^b$ & Model a & Model b & Model c\\
 & & $t_5$ (yr) & $t_5$ (yr) & $t_5$ (yr)
\\ \hline
dust-free & MRN & $4.9\times 10^{7}$ & $8.3\times 10^{6}$
& $7.3\times 10^5$\\
dusty & MRN & $3.3\times 10^{7}$ & $5.3\times 10^{6}$
& $4.5\times 10^5$ \\
dust-free & lognormal & $4.9\times 10^{7}$ & $8.3\times 10^{6}$
& $7.4\times 10^5$\\
dust-free & no dust$^c$ & $5.0\times 10^7$ & $8.6\times 10^6$
& $7.4\times 10^5$\\
\hline
\end{tabular}
\end{center}
$^a$The initial condition (IC) is chosen from the dust-free
and the dusty cases with the initial dust-to-gas ratio
$\mathcal{D}=0$ and 0.0075, respectively. For the dusty
initial condition, the grain
size distribution is assumed be the MRN (i.e.\
power law with an index of $-3.5$; see the text for details).\\
$^b$The grain size distribution of AGB dust is chosen from
MRN or lognormal.
$^c$This case is for no dust cooling (i.e.\ only gas cooling).
\end{minipage}
\end{table}

We compare the results with the different two initial conditions
for the dust-to-gas ratio, $\mathcal{D}_0=0$ and
0.0075 (dust-free and dusty initial conditions, respectively).
We adopt the MRN grain size distribution for AGB dust. As shown
in Fig.\ \ref{fig:tem_den},
the temperature drops more rapidly for the dusty initial
condition than for the dust-free initial condition
because of more efficient dust cooling. The time
at which the temperature drops down to $10^5$ K,
$t_5$, is
by a factor of $\sim$1.5 shorter for the dusty initial
condition than for the dust-free initial condition.
In Table \ref{tab:t5}, we list $t_5$ for each galaxy model
with different $\mathcal{D}_0$ and grain size
distributions of AGB dust.

In Fig.\ \ref{fig:dg}, we show the evolution of
dust-to-gas ratio, $\mathcal{D}$, calculated by
equation (\ref{eq:dg}). First we explain the evolution
of $\mathcal{D}$ for the dust-free initial condition.
At the beginning of the evolution, the dust-to-gas
ratio increases because of the dust supplied from the
AGB stars. The increase of $\mathcal{D}$ is saturated
because of the destruction by sputtering. At this stage,
the dust-to-gas ratio approaches to the value
determined by the balance between dust supply
and destruction as we show later.
Indeed, the sputtering time-scale given by equation (\ref{eq:tau_sp})
(note that $a<0.25~\micron$ for the MRN grain size distribution)
is much shorter than $t_5$, which allows the grain size distribution
to approach the equilibrium between supply and destruction.
At the end,
the dust-to-gas ratio decreases because the compression
of the gas due to the temperature drop by cooling
enhances the sputtering rate (but the temperature
is still higher than $\sim 2\times 10^6$~K). Indeed,
the decrease of dust-to-gas ratio coincides with the
temperature drop (Fig.\ \ref{fig:tem_den}).
As shown in equation (\ref{eq:sput}),
the sputtering efficiency scales as
$\propto\rho_\mathrm{gas}$ at any gas temperature,
while it is not
sensitive to the gas temperature as long as
$T\ga\tilde{T}=2\times 10^6$~K.
Because the effect of density is larger than that of
temperature, cooling eventually
enhances the dust destruction, contrary to the expectation
by \citet{mathews03}.

In order to show the level of $\mathcal{D}$ achieved
by the equilibrium between dust supply and destruction
under the initial gas density,
we also show the cases where we neglect
cooling (i.e.\ the gas temperature and density are constant)
in Fig.\ \ref{fig:dg}.
In these cases without cooling, the dust-to-gas ratio
is the same as the above case at the beginning of evolution
before the gas cools significantly, and it
approaches to the
value determined by the balance between dust supply and
sputtering. The dust-to-gas ratio achieved by the equilibrium
is the highest in Model~c, primarily because of the highest
stellar mass
per gas mass (i.e.\ the highest dust supply rate per gas mass).

We also present the results for the dust-rich initial
conditions ($\mathcal{D}_0=0.0075$) in Fig.\ \ref{fig:dg}a.
In Models a and b, the final
dust-to-gas ratio at $t_5$ is almost the same regardless
of $\mathcal{D}_0$ (note that $t_5$ is different, though, between
the different initial conditions; Table \ref{tab:t5}).
On the other hand, the effect of initial condition on the
final dust-to-gas ratio remains
in Model~c, because the evolution starts with lower temperature
so the effect of sputtering is smaller than Models a and b.
Yet, even in the dusty
initial condition of Model c, the dust-to-gas ratio
at $t=t_5$ is much lower than the initial dust-to-gas ratio
($\mathcal{D}_0=0.0075$).

Although the large fraction of the initial dust is destroyed in
any case, the initial condition has an imprint on the
temperature evolution as shown in Fig.\ \ref{fig:tem_den}.
Therefore, if the initial dust-to-gas ratio is as large as
the value in the Milky Way ISM, dust cooling has a
significant influence on the thermal evolution of the gas.
However, such a high dust-to-gas ratio is never achieved
if  the dust
enrichment and destruction take place simultaneously.

In order to show the effect of AGB dust, we also examine
the cases with and without dust enrichment for the dust-free
initial condition, finding
that dust cooling by AGB dust
has little influence on the
thermal evolution of the hot ISM.
Indeed, the evolutionary track is indistinguishable between
the cases with and without dust enrichment (thus, the
case without dust enrichment is not shown in Fig.\ \ref{fig:tem_den}).
This is also clear
from Table \ref{tab:t5} (compare the first and last
rows): $t_5$ is only
slightly longer with no dust cooling than with dust cooling.
We also confirmed that change of the grain size distribution
of AGB dust to the lognormal size distribution does not
alter the thermal history.
Indeed as shown in Table \ref{tab:t5}, $t_5$ is
almost identical between
the MRN and lognormal AGB dust.

In Fig.\ \ref{fig:dg}b, we examine the effects of
the grain size distribution of AGB dust on the evolution
of $\mathcal{D}$. Since the lognormal grain size distribution
is more biased to large sizes than the MRN grain size distribution,
it has a longer sputtering time-scale $\tau_\mathrm{sput}$
(equation \ref{eq:tau_sp}).
The final dust-to-gas ratio at $t\sim t_5$ in
the lognormal case is
about 4 times as large as in the MRN case.
Nevertheless, since the dust abundance is still kept much lower
than 0.01 by dust destruction,
the temperature evolution is not affected by dust cooling
even in the lognormal case as mentioned above.

\subsection{Grain size distribution}\label{subsec:size_distri}

\begin{figure}
\begin{center}
\includegraphics[width=0.4\textwidth]{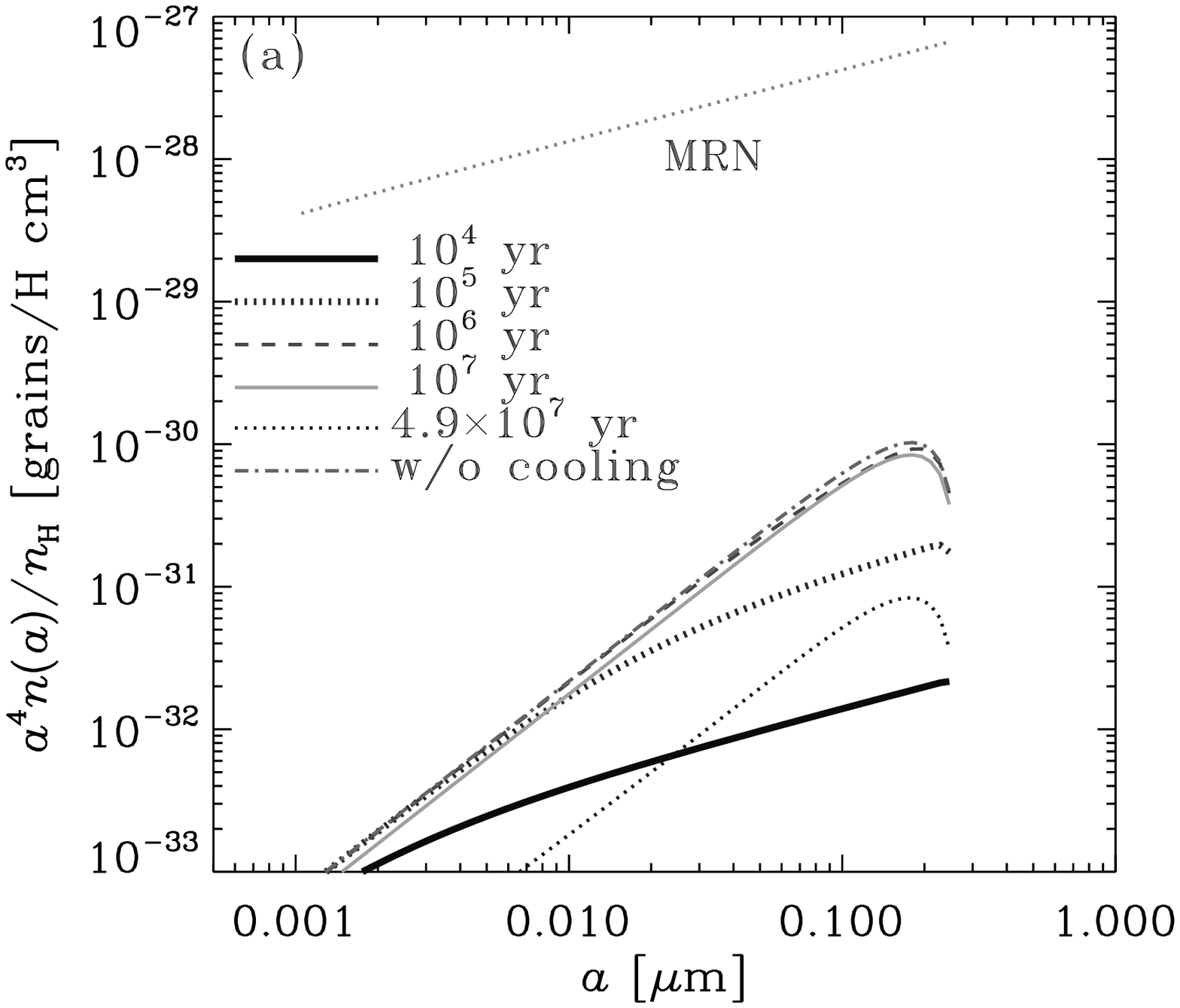}
\includegraphics[width=0.4\textwidth]{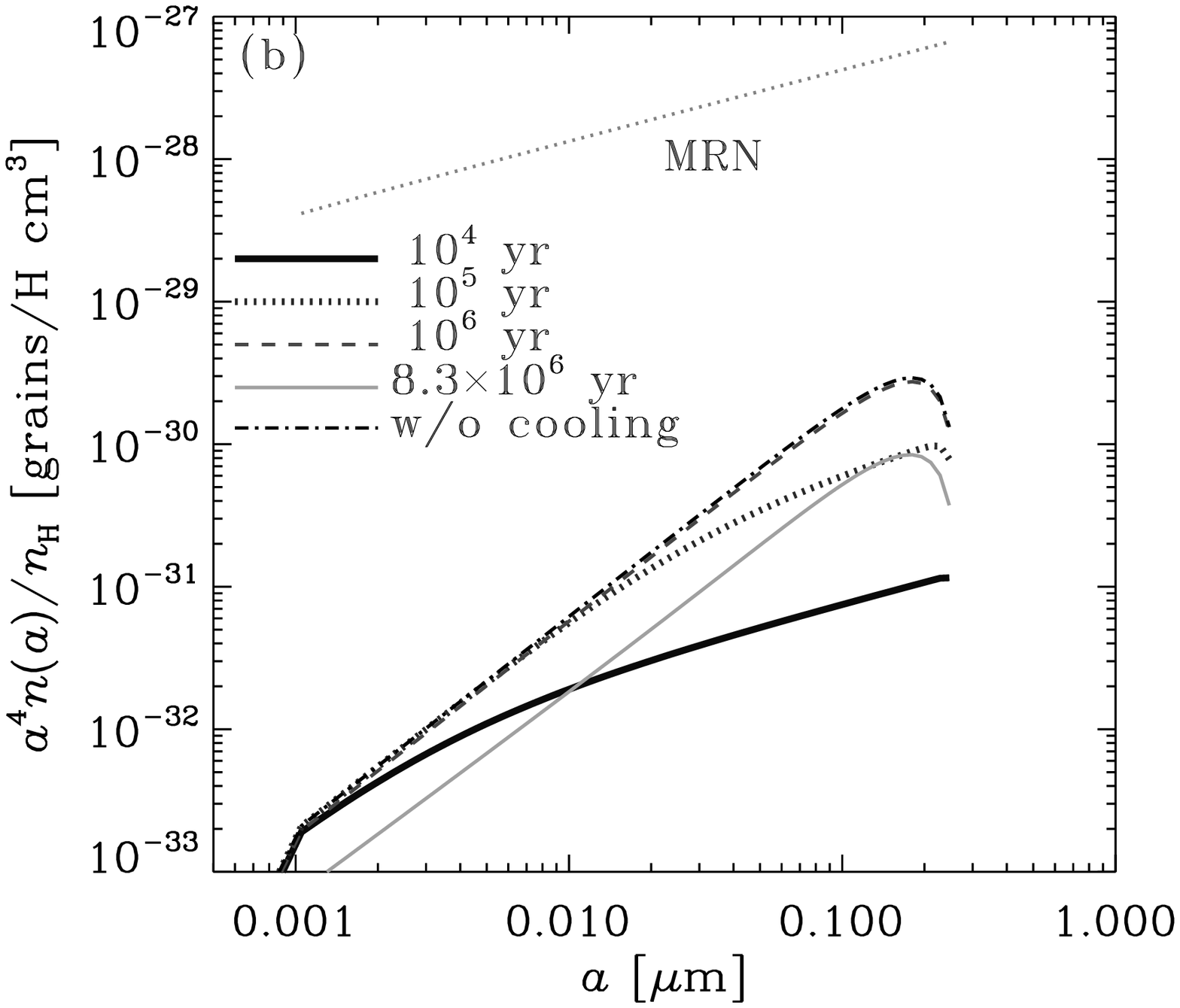}
\includegraphics[width=0.4\textwidth]{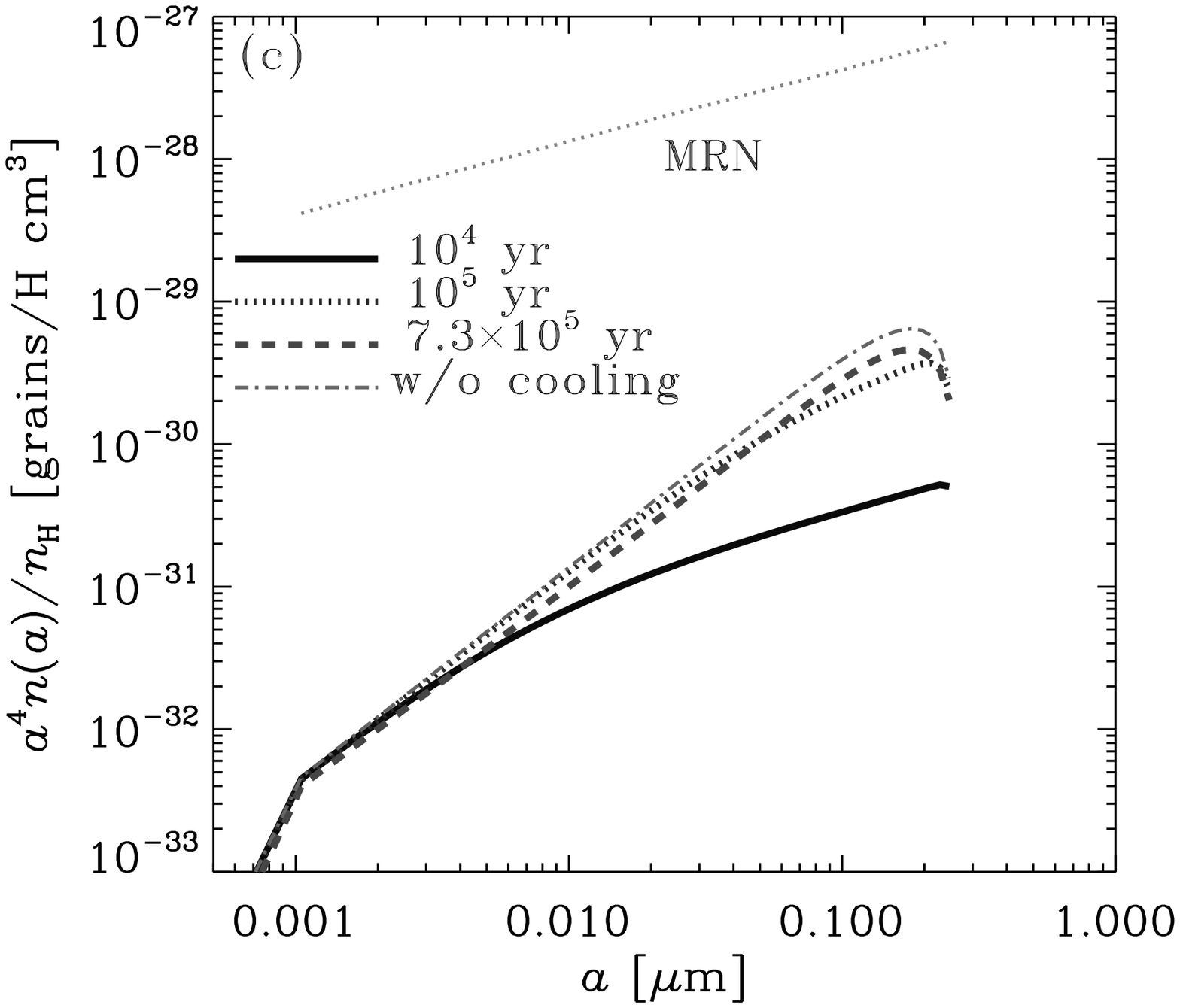}
\end{center}
\caption{Grain size distribution per hydrogen atom for
different times. The dust-free initial condition is adopted.
Grain size distributions are multiplied by $a^4$ to show
the grain mass distribution per logarithmic grain radius.
Panels (a), (b), and (c) show the results for Models a, b, and c,
respectively. The correspondence of the line species to the time
is shown in each panel. The dotted line marked as `MRN'
is the MRN grain size distribution with a dust-to-gas ratio
of 0.0075, shown as a reference for the grain size distribution
appropriate for the Milky Way ISM. Note that the final time
($t_5$) in each model is determined by the time when the gas
temperature reaches $T=10^5$~K. The dot-dashed line shows
the equilibrium grain size distribution without dust cooling.
\label{fig:size_distri}}
\end{figure}

Now we examine the evolution of grain size distribution in
detail.
In Fig.\ \ref{fig:size_distri}, we show the evolution of grain size
distribution for the dust-free initial condition.
The MRN grain size distribution is adopted for AGB dust.
In the initial stage at $t\ll t_5$,
when the change of gas density due to cooling is negligible,
the grain abundance gradually increases, converging to
the equilibrium grain size distribution determined by the
balance between the supply from AGB stars and the
destruction by sputtering, and the slope is described
by $n(a)\propto a^{-2.5}$ (Appendix~\ref{app:equilibrium}).
We have already seen above that the dust-to-gas ratio
approaches to the equilibrium value. As is also
observed above in Fig.\ \ref{fig:dg},
the grain abundance decreases at the final stage
of evolution around
$t\sim t_5$ because the density increase by cooling
enhances the sputtering rate. This is the reason why
the grain size distributions shift downwards at the
final stage in all the three models.

\begin{figure}
\begin{center}
\includegraphics[width=0.4\textwidth]{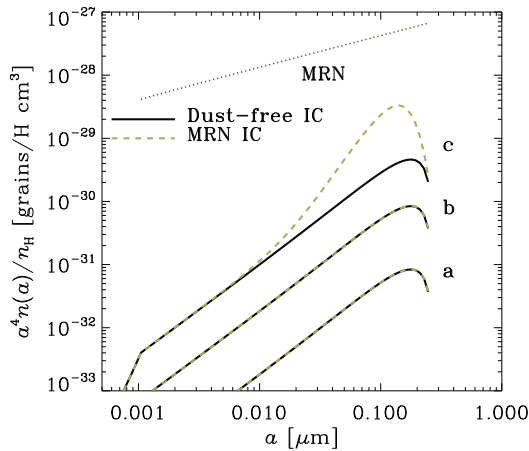}
\end{center}
\caption{Grain size distributions at $t=t_5$ (after cooling)
per hydrogen atom (multiplied by $a^4$).
Solid and dashed lines show the
results for the dust-free ($\mathcal{D}_0=0$) and
dust-rich ($\mathcal{D}_0=0.0075$) initial conditions
(IC), respectively.
The dotted line marked as `MRN'
is the MRN grain size distribution with a dust-to-gas ratio
of 0.0075, shown as a reference for the grain size distribution
appropriate for the Milky Way ISM.
\label{fig:size_distri_comp}}
\end{figure}

In Fig.\ \ref{fig:size_distri_comp}, we compare the
grain size distributions at $t=t_5$ for the three models
with the dust-free and dust-rich initial conditions
(the MRN grain size distribution is adopted for AGB dust).
As is consistent with Fig.\ \ref{fig:dg}, the dust
abundance per hydrogen
atom (or the dust-to-gas ratio) is the largest in Model~c:
as explained above, this is due to the highest stellar
mass per gas mass
(Table \ref{tab:param}), that is, the highest dust supply
rate per gas mass. Moreover, the comparison between
the cases of different $\mathcal{D}_0$ in Models a and b
shows almost no difference, which indicates that
almost all the dust that
existed initially in the hot gas has been destroyed. In Model c,
the effect of initial condition remains as explained in
Section \ref{subsec:tem} (Fig.\ \ref{fig:dg}): This is because of the low
temperature, i.e.\ low sputtering rate. The difference in the initial condition affects
at the largest grain sizes because the largest grains
have the longest sputtering time-scale (equation \ref{eq:tau_sp}).
However, even in Model c, the grain abundance
decreases far below the
initial MRN size distribution (shown by the dotted line in
Fig.\ \ref{fig:size_distri_comp}). Therefore, in all cases,
the final grain size distribution deviates significantly
from the initial grain size distribution.

\begin{figure}
\begin{center}
\includegraphics[width=0.4\textwidth]{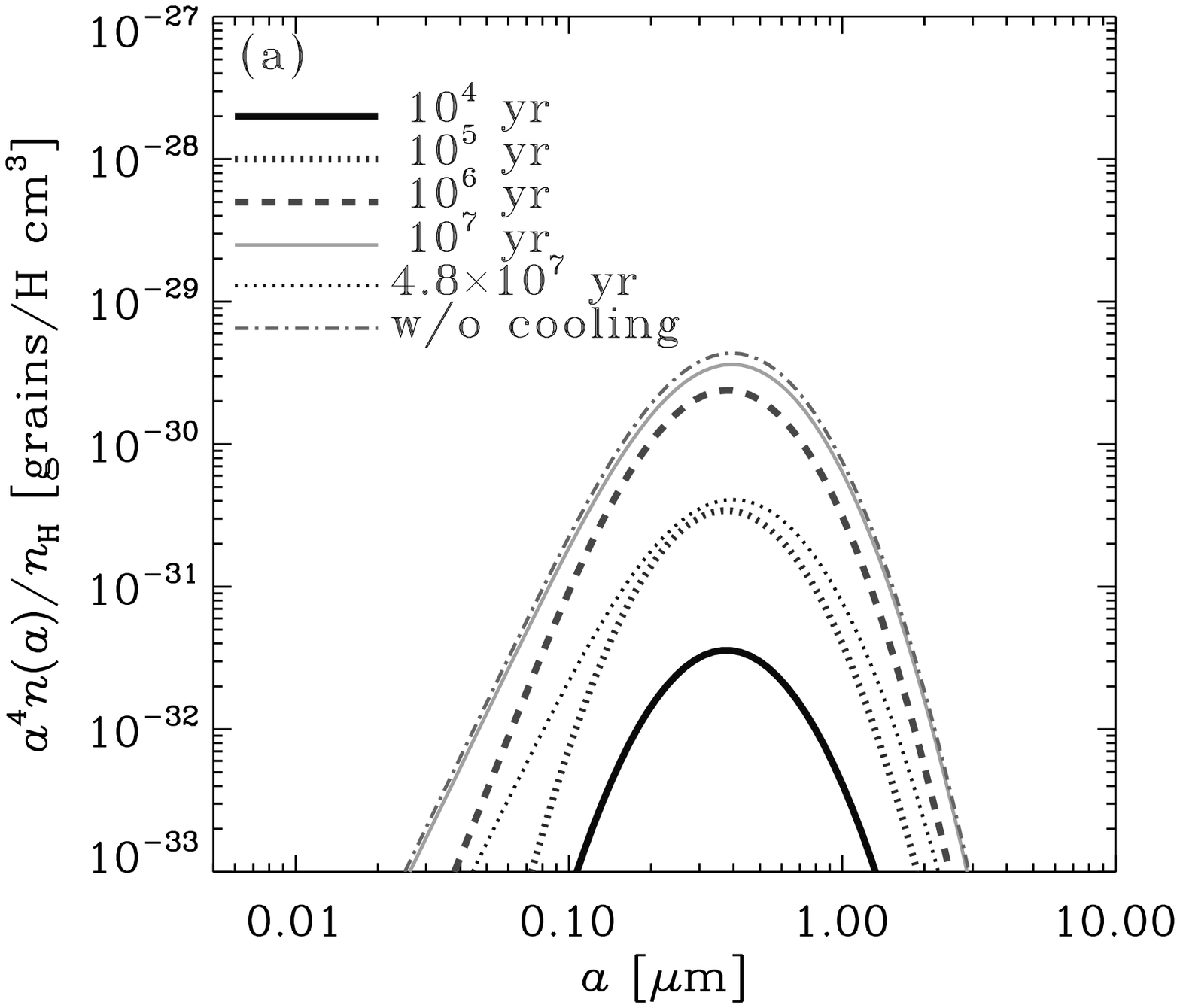}
\includegraphics[width=0.4\textwidth]{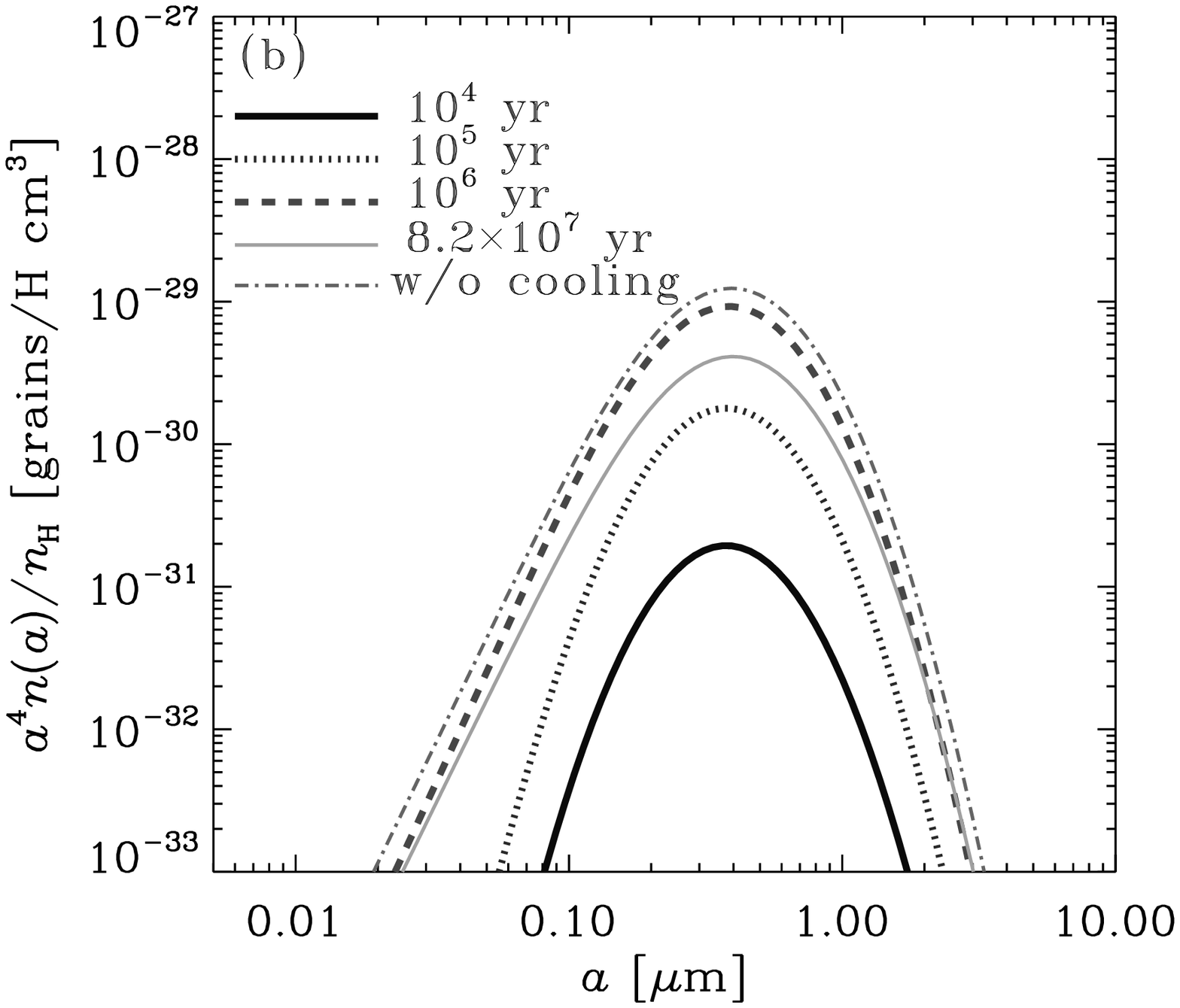}
\includegraphics[width=0.4\textwidth]{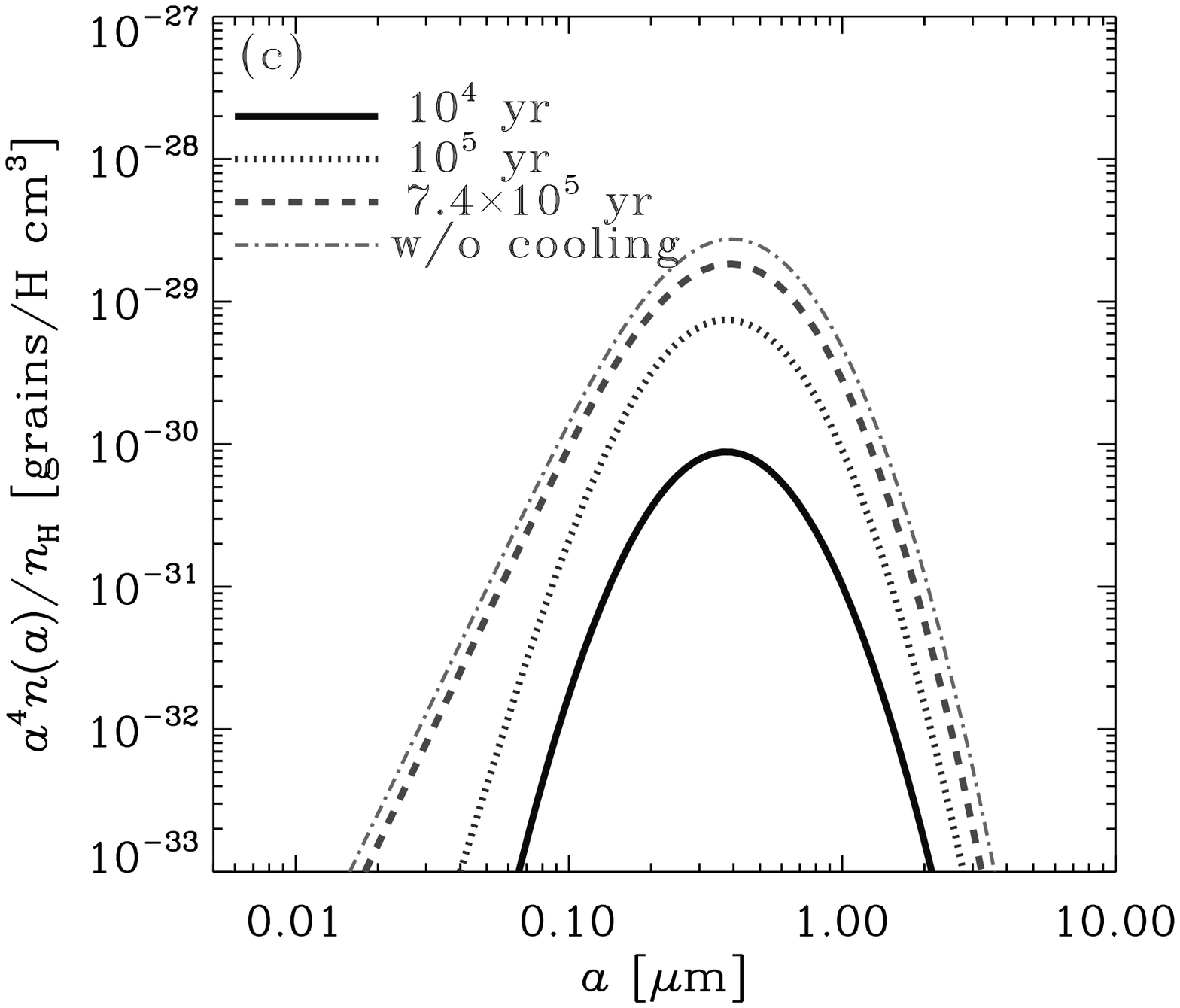}
\end{center}
\caption{Same as Fig.\ \ref{fig:size_distri} but for
the lognormal size distribution of AGB dust.
Panels (a), (b), and (c) show the results for Models a, b, and c,
respectively. The correspondence of the line species to
the time is shown in each panel. Note that the final time
($t_5$) in each model is determined by the time when the gas
temperature reaches $T=10^5$ K. The dot-dashed
line shows the equilibrium grain size distribution without
dust cooling.
\label{fig:size_distri_lognorm}}
\end{figure}

We also examine the case of the lognormal size distribution
for AGB dust (equation \ref{eq:lognorm}).
We adopt the dust-free initial condition.
The resulting grain size distributions are shown in
Fig.\ \ref{fig:size_distri_lognorm}, and their evolutionary
behaviour is similar to the
case of the MRN grain size distribution in the following
aspects: The grain abundance increases
before significant cooling occurs (i.e.\ at $t\ll t_5$),
converging to the equilibrium grain size distribution
(Appendix \ref{app:equilibrium}),
while it decreases after cooling (i.e.\ around
$t\sim t_5$). Compared with the lognormal
function, the tail toward small grain sizes becomes prominent
in the final grain size distributions at $t=t_5$. This tail is
due to the continuous destruction of large grains by
sputtering and described by $n(a)=\mbox{constant}$
(Appendix \ref{app:equilibrium}).
The grain abundance is reduced around $t=t_5$
because the increase of gas density by cooling enhances the
sputtering rate, as explained above.

\section{Processing after cooling}\label{sec:cool}

In the above, we have traced the dust evolution
in the hot gas up to the point where the gas cools down
to $10^5$ K. After that,
sputtering does not act as a dust processing mechanism any
more, but the dust may still be processed by other mechanisms
than sputtering. In particular, many authors have argued
the importance of dust growth by the accretion of
gas-phase metals
for the total dust budget in star-forming
galaxies \citep[e.g.][]{dwek98}. Since this accretion mechanism
works in cold and dense clouds, it is worth investigating
a possibility of dust growth also in the cold gas in
elliptical galaxies.
As mentioned in the Introduction (and shown
in Section~\ref{sec:result}), it has
been argued that the observed dust abundance in elliptical
galaxies cannot be explained by the internal dust supply
from AGB stars
because of rapid destruction by sputtering. Thus,
we examine dust growth as
a potential additional source of dust in elliptical galaxies
\citep[see also][]{fabian94,voit95,martini13}.

Under a high pressure in the inner part of elliptical
galaxy, a cold phase with $T\sim 10^2$ K is preferred
rather than a warm phase with $T\sim 10^4$ K \citep{wolfire95}.
Such a cold dense medium is indeed suitable for the site
of dust growth. Since accretion and coagulation
(grain--grain sticking) both occur
in such a cold dense medium, we treat them simultaneously
\citep{hirashita12}; however, as shown later, coagulation
only plays a minor role.
Below we calculate the dust evolution driven by accretion
and coagulation in the cold gas,
starting from the grain size distributions achieved after
cooling of the hot gas.

\subsection{Formulation of accretion and coagulation}

We calculate the evolution of grain size distribution by
accretion and coagulation using the formulation in
\citet{hirashita12}, which has a wide application
to, for example, polarization predictions
\citep{voshchinnikov14}. Again, the silicate dust properties
are assumed, but adopting
carbonaceous dust instead does not change the evolution of grain
size distribution significantly \citep{hirashita12}.
If we consider the
pressure equilibrium with the hot gas,
$n_\mathrm{H}> 10^3$ cm$^{-3}$ is achieved for all models
at a temperature appropriate for the cold ISM ($T\sim 10^2$ K). Thus,
we conservatively assume that $n_\mathrm{H}=10^3$ cm$^{-3}$,
but the results for other densities can be easily obtained
by noting that the time-scale of grain processing is
simply proportional to $n_\mathrm{H}^{-1}$.
The gas temperature, which determines the
thermal speed of the accreting material is assume to be
$10^2$ K.
The available gas-phase metals (in our case, silicon)
is estimated based on the
solar abundance of Si, $\mathrm{(Si/H)}=3.55\times 10^{-5}$
\citep{dappen00}.
We also assume that the mass fraction of Si in silicate is
0.166.
The grain velocity, used to estimate the grain--grain
collision rate for coagulation, is treated as a function
of grain radius,
and is taken from the molecular cloud case in \citet*{yan04},
which has a similar gas density to the one considered
here.
We also impose the same coagulation threshold velocity as
adopted in \citet{hirashita09} (originally based on
\citealt*{chokshi93,dominik97}):
only if the relative velocity is below this threshold,
which varies with the sizes of colliding grains,
the grains coagulate. However, detailed treatment of
coagulation does not affect the results,
because, as shown later, the abundant gas-phase metals
makes the role of accretion
stronger than that of coagulation in changing the grain
size distribution. For the initial condition
of each model, we use the grain size distribution
achieved in the hot gas after cooling. More specifically,
the initial
grain size distribution is produced by multiplying
the grain size distributions per hydrogen
at $t=t_5$, $n(a,\, t_5)/n_\mathrm{H}$, in Section
\ref{sec:result} by $n_\mathrm{H}=10^3$ cm$^{-3}$.
The time is newly measured from the onset of grain growth
in the cold gas; to avoid the confusion, we use a different
notation for the time, $t'$, measured from the onset of
grain growth.

\subsection{Results}\label{subsec:result_growth}

\begin{figure}
\begin{center}
\includegraphics[width=0.45\textwidth]{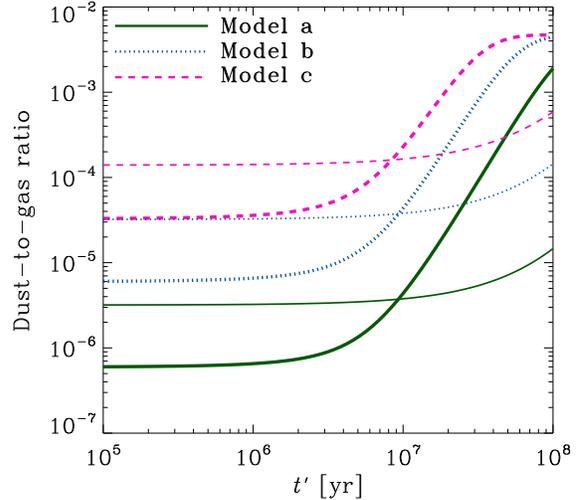}
\end{center}
\caption{Time evolution of dust-to-gas ratio $\mathcal{D}$
by dust growth. The time $t'$ is the elapsed time from the
onset of dust growth in cooled gas. The thick solid, dotted,
and dashed lines present the results for Models a, b, and
c in the case where we use the result of the MRN AGB dust
calculation at $t_5$ for the initial condition (i.e.\ the
final state of each model
in Fig.\ \ref{fig:size_distri}
is used for the initial condition). The thin lines with the
same line species represent the results for the lognormal
AGB dust calculation at $t_5$ (i.e.\ the final state of
each model in Fig.\ \ref{fig:size_distri_lognorm} is used
for the initial condition).
\label{fig:dg_growth}}
\end{figure}

First, we show in Fig.\ \ref{fig:dg_growth}
the evolution of dust-to-gas ratio
$\mathcal{D}$ estimated by equation (\ref{eq:dg})
at $t'\leq 10^8$ yr
(as shown later, the observed
extinction curves are well covered by the grain
size distributions at $t'\leq 10^8$ yr).
Note that accretion changes $\mathcal{D}$ while
coagulation does not.
If we start with the grain size distribution
at $t_5$ for the MRN AGB dust, $\mathcal{D}$
begins to increase drastically from
$t'\sim\mbox{a few}\times 10^6$ yr and
reaches $10^{-3}$ in a few $\times 10^7$ yr.
Therefore, if the lifetime of the cold gas is longer than
$10^7$ yr, there is a possibility of dust `reformation'
using the remnants of sputtered dust grains,
which explains the observed level of dust abundance
as we discuss in the next section.
The increase of $\mathcal{D}$ saturates most quickly
in Model c simply because the initial dust abundance
is the largest, which also means that the abundance of
available gas-phase metals is the lowest.

For the calculation adopting the lognormal AGB
dust case for the initial condition, as shown
in Fig.\ \ref{fig:dg_growth},
the increase of
dust-to-gas ratio is not so drastic as in the case of
MRN AGB dust models
because the grain size distribution is biased to
large sizes (the accretion time-scale is proportional
to the grain radius; \citealt{hirashita12}).
This indicates that the initial
grain size distribution in the cold gas, which is the
final grain size distribution of the cooled hot gas,
is of fundamental importance in determining the efficiency
of dust mass increase in the cold gas.

\begin{figure}
\begin{center}
\includegraphics[width=0.4\textwidth]{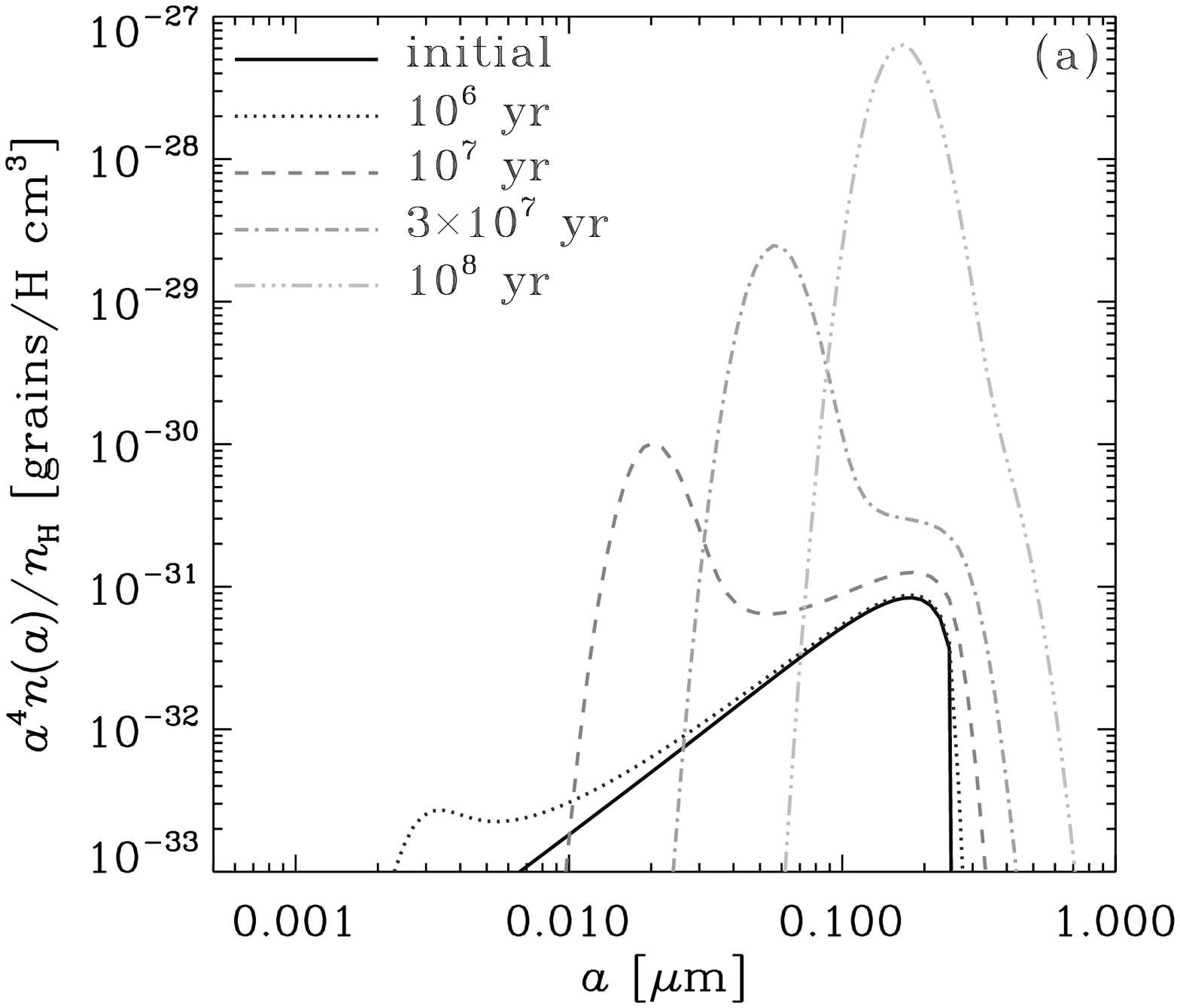}
\includegraphics[width=0.4\textwidth]{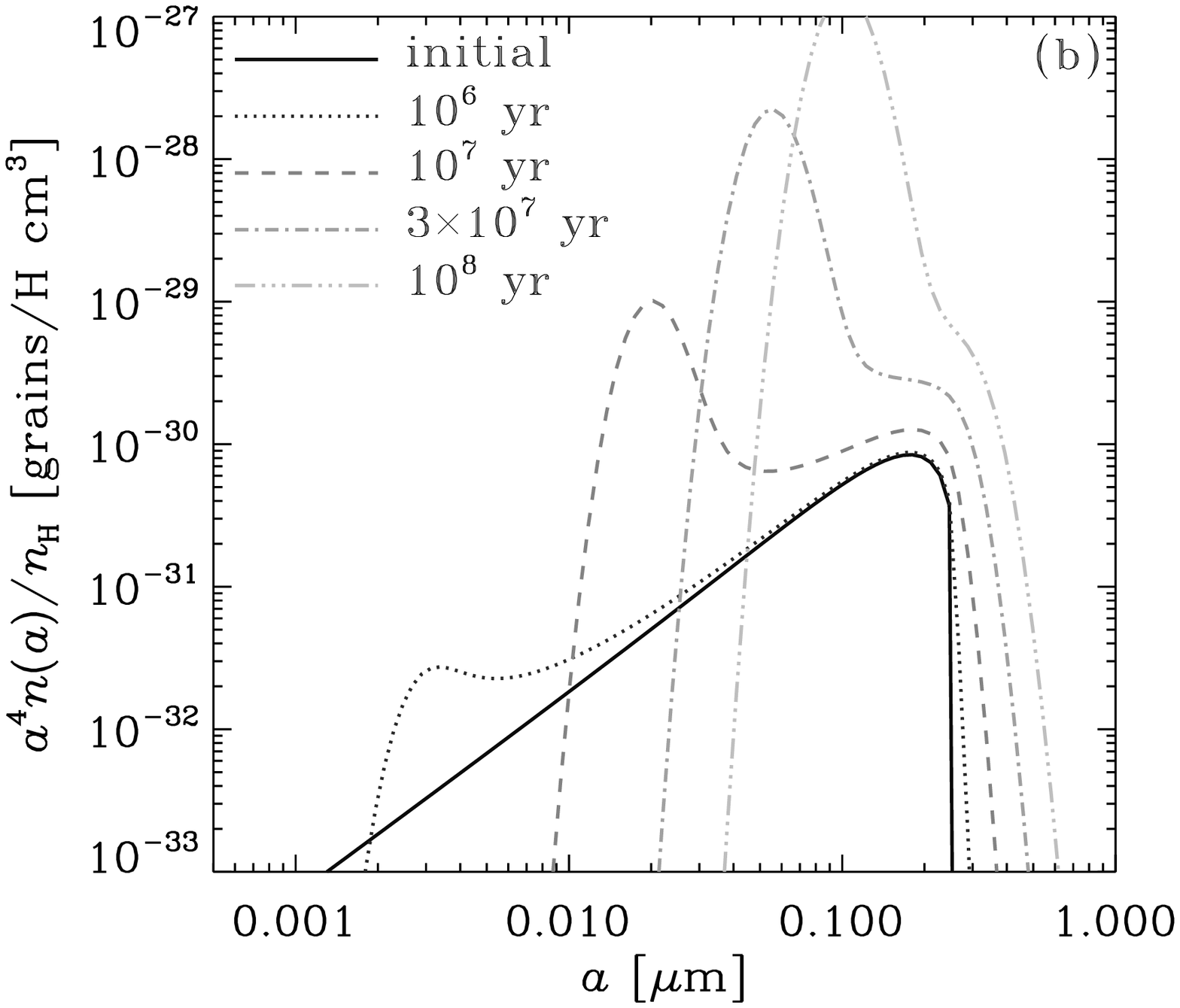}
\includegraphics[width=0.4\textwidth]{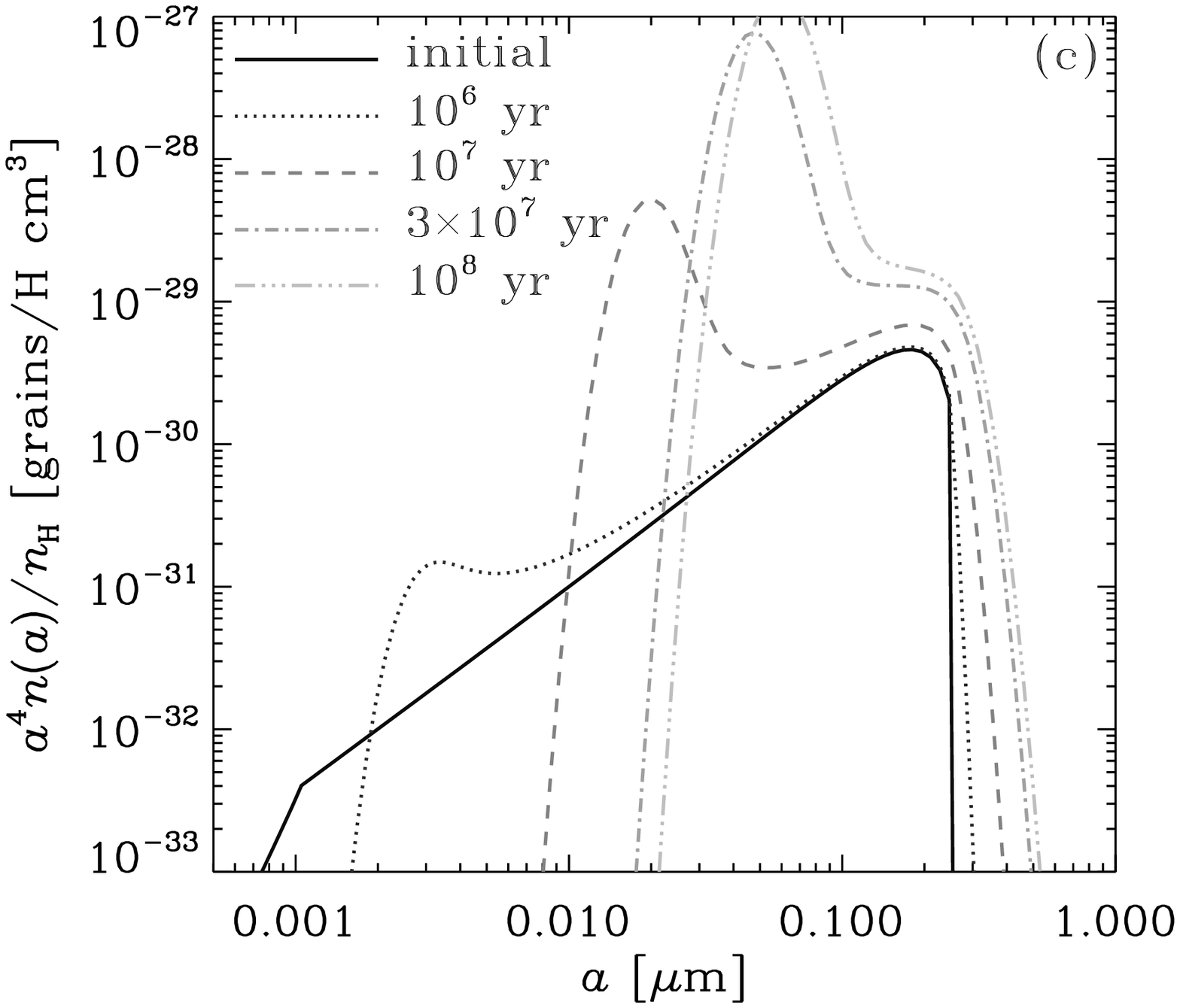}
\end{center}
\caption{Evolution of grain size distribution by grain
growth, starting from the final grain size distributions in
Fig.\ \ref{fig:size_distri}.
Panels (a), (b), and (c) show the results for Models a, b, and c,
respectively. The solid, dotted, dashed, dot-dashed,
dot-dot-dot-dashed lines present the grain size distributions
at $t'=0$, $10^6$, $10^7$, $3\times 10^7$ and $10^8$ yr,
respectively.
We assume $n_\mathrm{H}=10^3$ cm$^{-3}$, but the
time-scale just scales as $\propto n_\mathrm{H}^{-1}$.
\label{fig:growth}}
\end{figure}

Next, we investigate the evolution of grain size distribution
by accretion and coagulation.
In Fig.\ \ref{fig:growth}, we show the evolution of grain
size distribution, adopting the MRN AGB dust case for
the initial condition. We observe
that accretion continue to increase the dust abundance
over $\sim 10^8$ yr, which is consistent with the evolution
of $\mathcal{D}$ shown above. Because accretion saturates
most quickly in Model c as explained above, the grain radii
achieved by the growth are the smallest in Model c.

The most important feature of grain growth by accretion
is that the effect appears most significantly at the
smallest grain sizes.
As explained in \citet{hirashita12}, accretion
enhances the dust abundance at the smallest grain sizes,
which have the shortest accretion time-scale because of
the largest grain surface-to-volume ratio.
In order to isolate the effect of accretion,
we also examined the grain size distribution with only
accretion (i.e.\ without coagulation). We confirmed that there
is almost no difference between the cases with and without
coagulation (we do not plot the results because
the grain size distributions with and without
coagulation are almost indistinguishable).
This is partly because the effect of accretion
is enhanced by the large amount of available
gas-phase metals, partly because coagulation does not
proceed beyond $a\sim 0.1~\micron$ at which grains have
larger velocities than the coagulation threshold
(this means that the creation of larger grains than
$\sim 0.1~\micron$ only occurs by accretion).

\begin{figure}
\begin{center}
\includegraphics[width=0.4\textwidth]{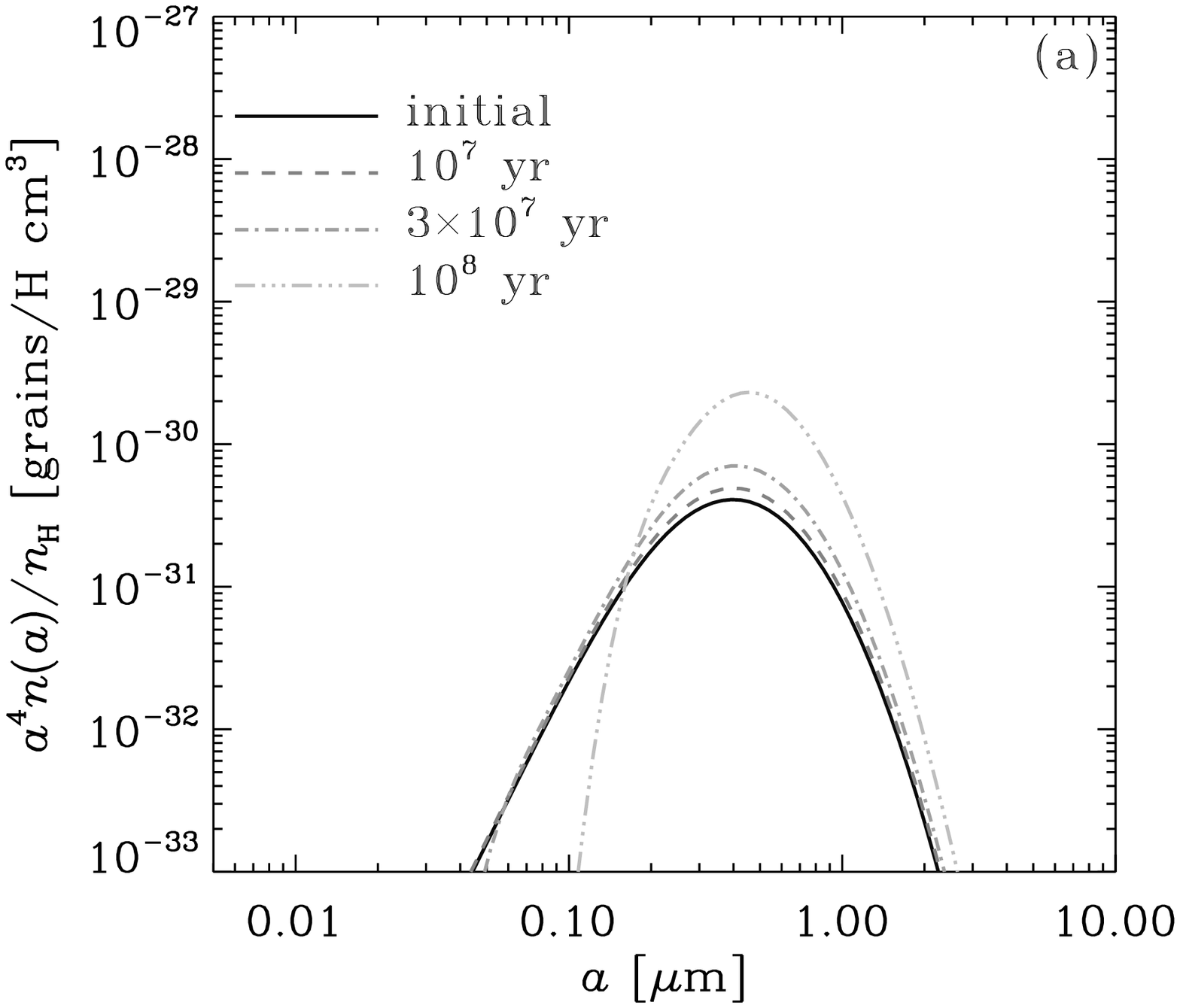}
\includegraphics[width=0.4\textwidth]{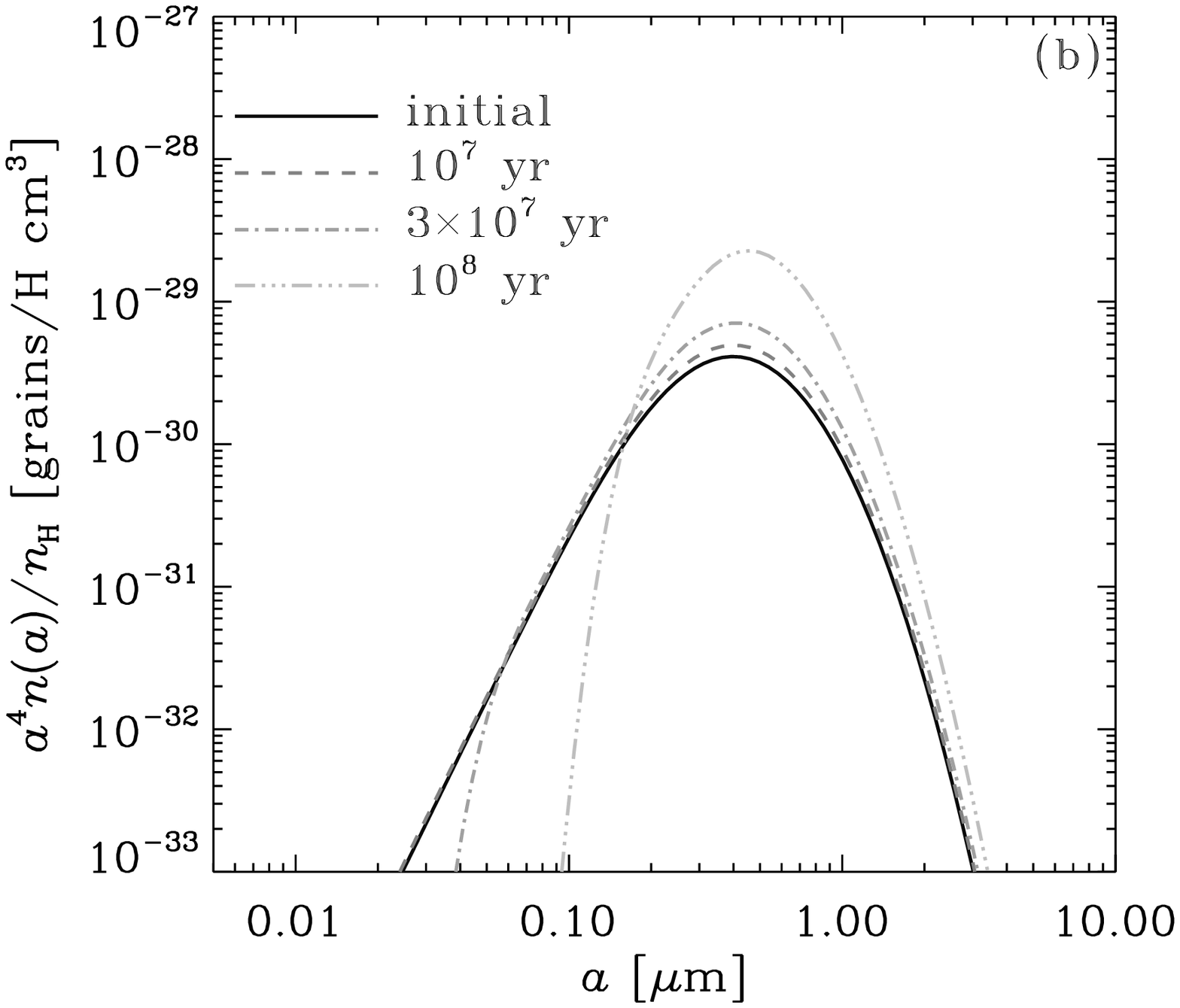}
\includegraphics[width=0.4\textwidth]{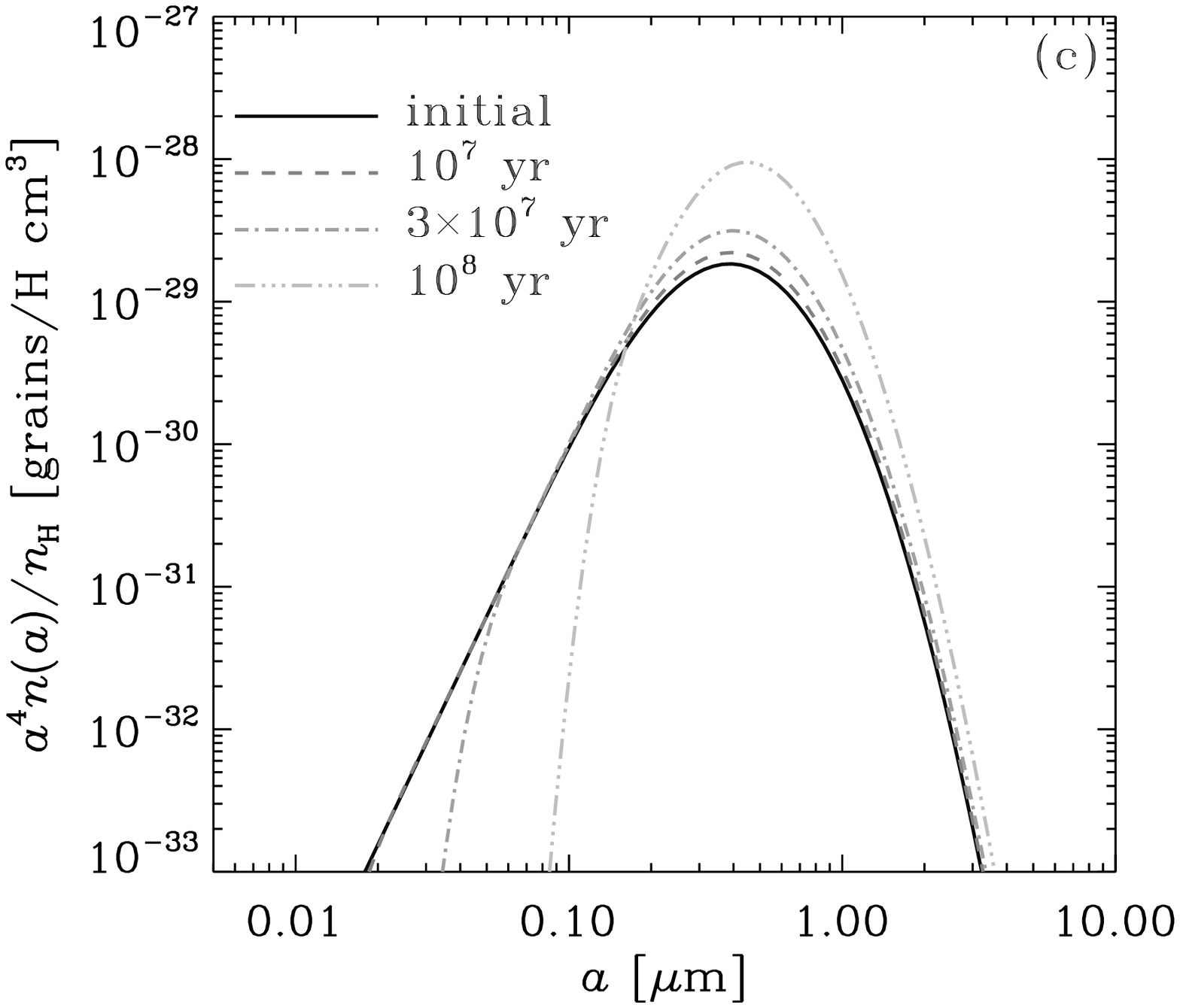}
\end{center}
\caption{Evolution of grain size distribution by grain
growth, starting from the final grain size distributions in
Fig.\ \ref{fig:size_distri_lognorm}.
Panels (a), (b), and (c) show the results for Models a, b, and c,
respectively. The solid, dashed, dot-dashed,
dot-dot-dot-dashed lines present the grain size distributions
at $t'=0$, $10^7$, $3\times 10^7$ and $10^8$ yr, respectively.
We assume $n_\mathrm{H}=10^3$ cm$^{-3}$, but the
time-scale just scales as $\propto n_\mathrm{H}^{-1}$.
\label{fig:growth_lognorm}}
\end{figure}

In Fig.\ \ref{fig:growth_lognorm}, we show the same plot
as above but we adopt the lognormal AGB dust case
(i.e.\ the final grain size distribution in
each panel of Fig.\ \ref{fig:size_distri_lognorm}) for the
initial condition. We observe that the change of
the grain size distribution occurs more slowly than
the MRN AGB dust case because the grains are biased to
larger sizes (recall that the accretion time-scale is
proportional to the grain radius). Therefore, grain growth is
very sensitive to the size distribution of
the grains incorporated in the cold medium.
However, as shown later, extinction curves calculated
for the lognormal AGB dust cases
are too flat and are inconsistent with the observed
extinction curves (Section \ref{subsec:ext}).

\section{Observational properties and Discussion}\label{sec:obs}

\subsection{Dust abundance}

For comparison of dust abundance, we adopt
elliptical galaxy samples observed by
\textit{Herschel} from \citet{smith12} and
\citet{alighieri13}. The dust-to-gas ratio is
estimated for each galaxy
based on the dust mass derived from the
\textit{Herschel} FIR emission and the gas mass
estimated by the sum of H \textsc{i} and H$_2$ masses
(the latter is from CO luminosity). The sample
galaxies adopted are listed in Table \ref{tab:sample}.
For the sample in \citet{smith12},
we select 4 galaxies with
(i) a morphological type of elliptical galaxies (E); (ii) detection of
FIR emission (i.e.\ a successful estimate of dust mass); and
(iii) detection of H \textsc{i} mass.
We use the upper limits for H$_2$ mass, but the treatment of
H$_2$ mass does not affect the results since these upper limits
of H$_2$ mass are significantly smaller than the H \textsc{i} mass.
For the sample in \citet{alighieri13}, we exclude galaxies with
morphological type of S0 and Sa and only select the objects
for which both H \textsc{i} and H$_2$ masses are constrained
(i.e.\ detected or with an upper limit obtained). As a consequence,
we adopt
4 objects from their sample.
NGC 4374 is common to both samples.
The $B$-band and X-ray luminosities are taken from
\citet*{osullivan01}.

\begin{table*}
\centering
\begin{minipage}{140mm}
\caption{Elliptical galaxy sample adopted from \citet{smith12}
and \citet{alighieri13}.}
\label{tab:sample}
\begin{center}
\begin{tabular}{@{}llccccccc} \hline
Name & Other name & $\log L_B$ & $\log L_\mathrm{X}$ & $\log M_\mathrm{dust}$ &
$\log M_\mathrm{HI}$ & $\log M_\mathrm{H_2}$ & $\log\mathcal{D}$ & Ref.$^a$\\
 & & (L$_{\sun}$) & (erg\,s$^{-1}$) & (M$_{\sun}$) & (M$_{\sun}$) &
(M$_{\sun}$) & \\ \hline
VCC 763 & NGC 4374, M84 & 10.57 & 40.83 & 5.05$^b$ & 8.96 & $<$7.23 &
$-$3.91$^c$ & 1, 2, 3\\
HRS 150 & NGC 4406, M86 & 10.66 & 42.05 & 6.63 & 7.95 & $<$7.4 &
$-$1.43$^c$ & 1, 3\\
HRS 186 & NGC 4494 & 10.62 & $<$40.10 & 5.08 & 8.26 & $<$7.35 &
$-$3.23$^c$ & 1, 3\\
HRS 241 & NGC 4636 & 10.51 & 41.59 & 5.06 & 9.0 & $<$7.02 &
$-$3.94$^c$ & 1, 3 \\
VCC 345 & NGC 4261 & 10.70 & 41.21 & 5.81 & $<$8.45 & $<$7.70 &
$>-$2.71$^d$ & 2, 3\\
VCC 1226 & NGC 4472, M49 & 10.90 & 41.43 & 5.49 & $<$7.90 & $<$7.26 &
$>-$2.50$^d$ & 2, 3\\
VCC 1619 & NGC 4550 & 9.72 & 39.78 & 5.41 & $<$7.90 & 7.20 &
$>-$2.57$^d$ & 2, 3\\
\hline
\end{tabular}
\end{center}
$^a$References: 1) \citet{smith12}; 2) \citet{alighieri13}; 3) \citet{osullivan01}.\\
$^b$Adopted from Ref.\ 1. Ref.\ 2 gives 5.30 with the same mass
absorption coefficient.\\
$^c$The upper limit of H$_2$ mass is used, but this does not
affect the estimated dust-to-gas ratio since the gas mass is
dominated by H \textsc{i} gas mass.\\
$^d$We use the upper limits of gas mass to derive the lower
limits of dust-to-gas ratio.
\end{minipage}
\end{table*}

As seen in Table \ref{tab:sample}, the observed
dust-to-gas ratio is larger than $10^{-4}$.
However, as shown in Fig.~\ref{fig:dg}, this level of
dust-to-gas ratio cannot be achieved in the hot ISM,
especially in Models a and b, which are appropriate for
most of the $B$-band luminosities of the sample.
The situation does not change even if we change the
grain size distribution of AGB dust to the lognormal
distribution (Fig.~\ref{fig:dg}b).
This confirms previous conclusions derived by various
authors that the dust destruction in the hot gas phase
prevents the dust-to-gas ratio (or the total dust mass)
from reaching
the observed level \citep{goudfrooij95,patil07}, and
shows that this is also true even if we consider the grain
size distribution. We have also shown
that cooling does not help the dust to survive, contrary to
\citet{mathews03}'s expectation: Gas
compression caused by cooling enhances the
sputtering efficiency, leading to an additional decrease of
the dust abundance. In other words, the
effect of cooling further emphasizes the discrepancy
between the expected and observed dust abundances
in elliptical galaxies.

Some authors argue an external origin of the gas and
dust based on the evidence that the kinematic
properties of the gas are decoupled from those of the major stellar
component in some elliptical galaxies \citep{forbes91,caon00}.
Since not all the galaxies fit this external-origin
scenario \citep{forbes91},
it is still worth considering how much dust could be
explained by the `internal-origin' scenario in which the dust is
produced by their own stellar populations and/or inherent
dust growth mechanisms.

Regarding the internal origin of dust, we have shown that
the accretion of gas-phase metals in cold clouds increases the
dust-to-gas ratio to a level of $\sim 10^{-3}$ in
$\sim 10^7$--$10^8$ yr. In this sense,
cooling is important in the dust abundance, not through
survival in the hot gas (as in \citealt{mathews03}'s
original idea) but through the formation of cold clouds
hosting dust growth.
The time-scale of the dust abundance increase by dust growth
is shorter than the feedback time-scale
($\ga 10^8$ yr; \citealt{kaviraj11,pellegrini12}),
on which AGN activity heats the cooled gas.
Therefore, the `recovery' of dust by accretion is
a plausible scenario of explaining the observed excessive
dust abundance in elliptical galaxies.

\subsection{Extinction curves}\label{subsec:ext}

The grain size distributions obtained by our calculations
are tested against the observed extinction curves in \citet{patil07}.
Based on the calculated grain size distributions above,
we theoretically predict the corresponding extinction curves.
The extinction at wavelength $\lambda$ in units of magnitude
($A_\lambda$) is calculated by weighting the extinction
efficiency factor $Q_\mathrm{ext}(a,\,\lambda )$ with the
grain size distribution $n(a)$ as
\begin{eqnarray}
A_\lambda =C\int_0^\infty n(a)\upi a^2Q_\mathrm{ext}(a,\,\lambda )
\mathrm{d}a,
\end{eqnarray}
where $C$ is the normalizing constant, which cancels out
when we consider the ratio of $A_\lambda$ at two wavelengths
later. Although we assumed silicate for the grain property,
the grain growth time-scale of carbonaceous dust is similar
\citep{hirashita12}. The steepness of extinction curve is
often discussed in comparison with the Milky Way extinction curve.
Therefore, we simply assume a mass fraction of
0.54 : 0.46 for silicate and carbonaceous dust (graphite),
which fits the
Milky Way extinction curve under the MRN grain size distribution
\citep{hirashita09}. We refer to \citep{hirashita09} for the
adopted parameters
for silicate and graphite.
However, since we are interested
in wavelengths longer than 0.4~$\micron$,
the extinction curve slope is broadly determined by the
grain size distribution with a minor effect of the material.

\begin{figure}
\includegraphics[width=0.45\textwidth]{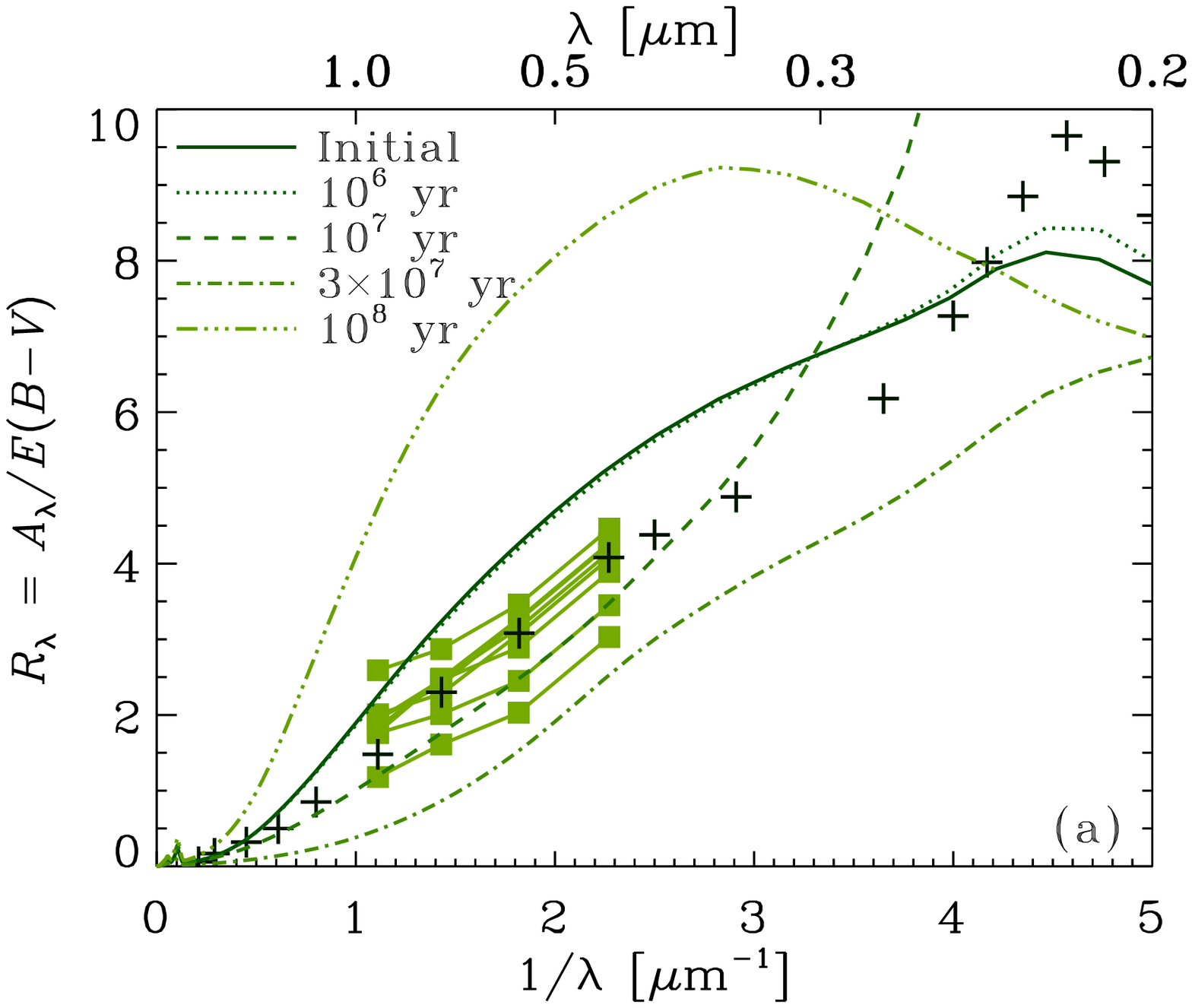}
\includegraphics[width=0.45\textwidth]{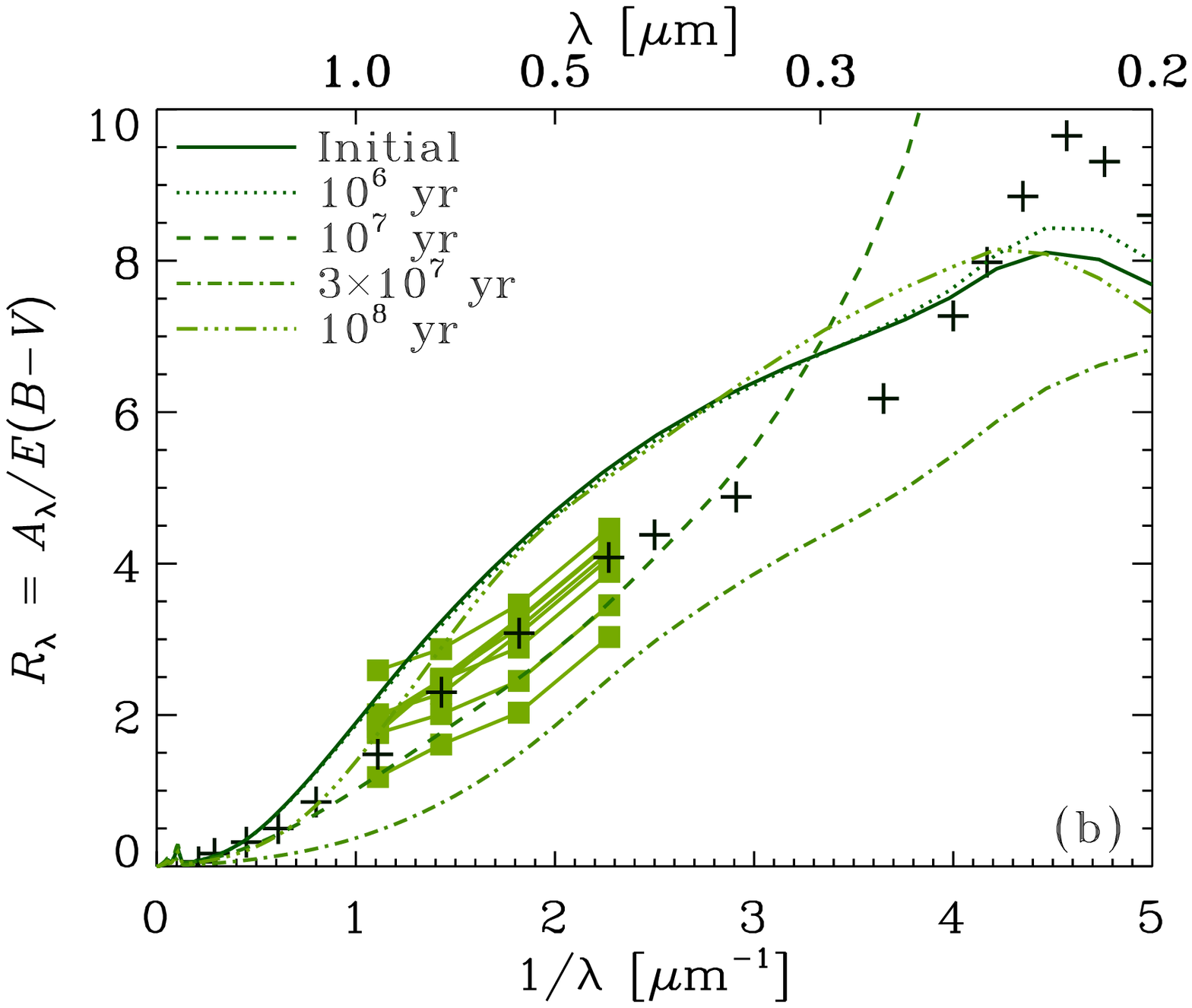}
\includegraphics[width=0.45\textwidth]{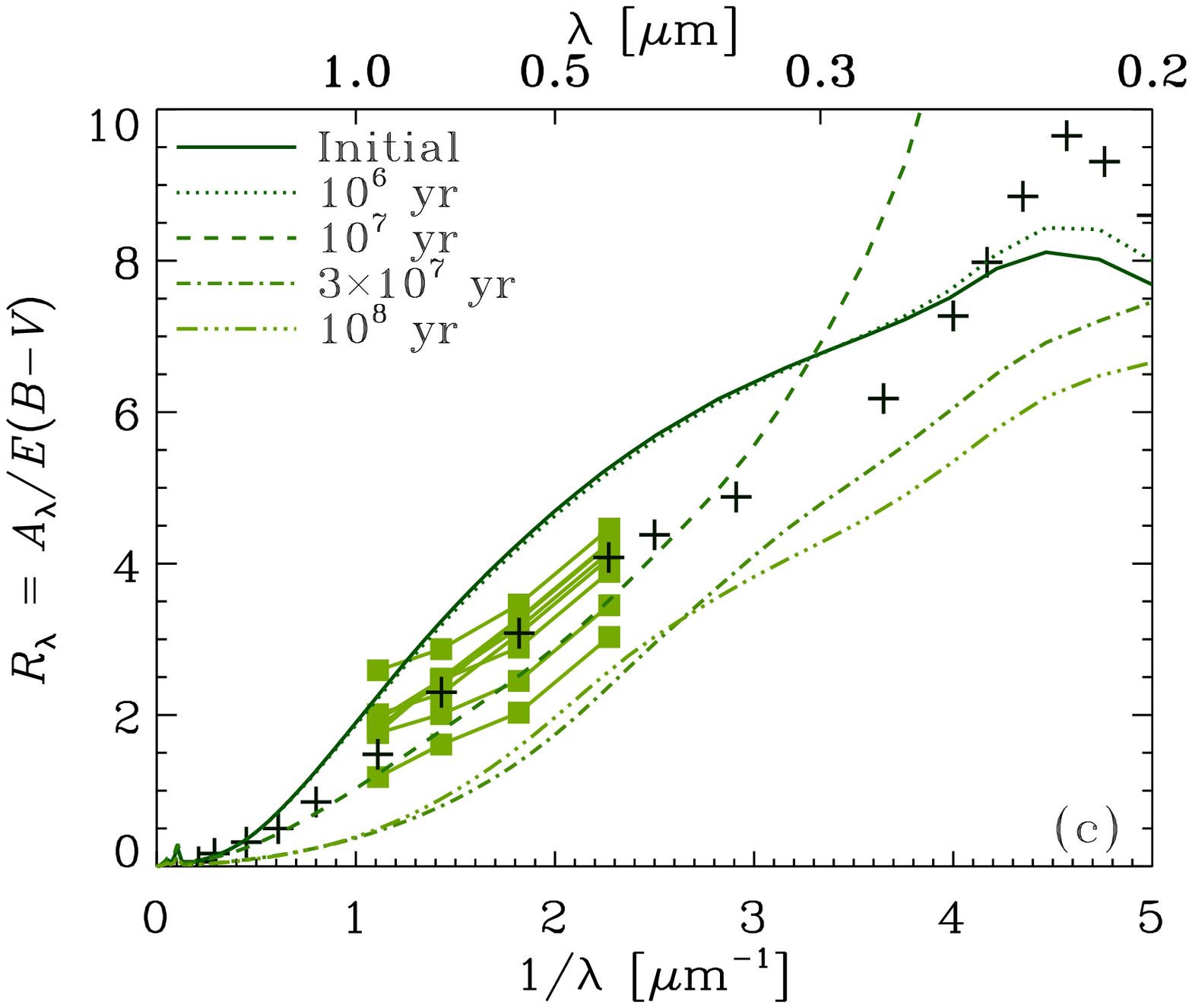}
\caption{Extinction curves normalized to $E(B-V)$.
Panels (a), (b), and (c) show
the results for Models a, b, and c, respectively.
The solid curve shows the extinction curves at the final
stage of gas cooling in Fig.\ \ref{fig:size_distri},
which corresponds to the extinction curve at the onset of
grain growth in the cold medium. The dotted, dashed,
dot-dashed, dot-dot-dot lines present the extinction curves
at $10^6$, $10^7$, $3\times 10^7$, and $10^8$ yr after the
onset of grain growth. The squares connected by the solid line
are observational data for an elliptical galaxy sample in
\citet{patil07}. The crosses show the Milky Way extinction
curve as a reference.
\label{fig:extinc}}
\end{figure}

Now we examine the extinction curves in the MRN AGB dust case.
In Fig.\ \ref{fig:extinc}, we show the extinction curves
for the grain size distributions at $t_5$
in Fig.\ \ref{fig:size_distri}
(i.e.\ the initial grain size distributions before grain growth)
and the variation of extinction curves by
grain growth (time is measured from the onset of grain growth).
We show the total to selective extinction
$R_\lambda\equiv A_\lambda /E(B-V)=A_\lambda/(A_B-A_V)$
(the wavelengths at the $B$ and $V$ bands are 0.44 and 0.55
$\micron$, respectively), which is sensitive to the
extinction curve slope around these two optical bands.
The observational extinction curves for a sample of elliptical
galaxies are taken from \citet{patil07}: we only chose those
galaxies that are classified as elliptical galaxies and
have extinction data at all bands ($B$, $V$, $R$, and $I$).

We observe in Fig.\ \ref{fig:extinc} that the
initial extinction curves before grain growth tend to
overproduce $R_\lambda$ since they are too flat [i.e.\ the
colour excess
$E(B-V)$ is small] compared with the observed extinction curves.
They are also flatter than the
Milky Way extinction curve, which has $R_V\simeq 3.1$
\citep[e.g.][]{draine03}, because of the deficiency of
small grains (note that the grain size distribution
after sputtering in Fig.\ \ref{fig:size_distri} is
$\propto a^{-2.5}$ while the Milky Way extinction curve
is fitted by a grain size distribution $\propto a^{-3.5}$;
MRN). After grain growth by accretion takes place,
the extinction curve
becomes steeper (or $R_\lambda$ becomes smaller), which
is consistent with the previous conclusion that grain
growth by accretion steepens the extinction curve
\citep{hirashita12}.
This is because, as explained in
Section \ref{subsec:result_growth}, accretion is
much more efficient for
smaller grains owing to their larger surface-to-volume
ratio. Since $R_\lambda$ is sensitive to
the slope at $B$ and $V$ bands, the decrease of
$R_\lambda$ is significant if the abundance of grains with
$a\la\lambda /2\upi\simeq 0.07~\micron$ (for
$\lambda =0.44~\micron$) increases \citep{bohren83}.
This occurs up to $3\times 10^7$ yr in Model a
and $10^8$ yr in Models b and c (Fig.~\ref{fig:growth}).
At $10^8$ yr in
Model a, almost all grains exceed 0.07 $\micron$, leading
to a significant increase of $R_\lambda$ (or significant
flattening of optical extinction curve), as seen in
Fig.~\ref{fig:extinc}a.

It is not likely that we observe the extinction of single-aged
clouds, but it is probable that we see a mixture of clouds with different
ages at the same time. Considering this, it is interesting to point out
that the observed extinction curves are just in the range
explained by the grain growth. Sputtering in
the hot gas tends to make the extinction
curve too flat; the observed extinction curves favour
subsequent grain processing that could form grains
with $a\la 0.07~\micron$. Grain growth by accretion
enhances the abundance of such `small' grains.

The lognormal grain size distribution of AGB dust produces
completely flat extinction curve at $\lambda\la 1~\micron$,
even if we consider accretion. Since $E(B-V)$ is slightly
negative (i.e.\ $R_\lambda$ is negative and $|R_\lambda|$
is extremely large), the results cannot be displayed.
Therefore, the lognormal model
does not fit the observed extinction curve unless there
is an extra mechanism of small grain production
(see Section \ref{subsec:shat}).

\subsection{Shattering?}\label{subsec:shat}

\citet{hirashita09} also showed that the dust grains may be
shattered in the warm ($\sim 10^4$ K) diffuse
medium in the Galactic environment \citep[see also][]{yan04}.
If some fraction of the gas remains to be diffuse after
cooling of the hot gas
and turbulent velocity is as high as $\ga 1$ km s$^{-1}$,
grains may be shattered. This would enhance the abundance of
small grains and would steepen the extinction curve.
If the dust is processed by shattering before being included
in the cold dense medium, the total surface area of the dust
is enhanced, resulting in more efficient dust growth by
accretion and further steepening of the extinction curve.
Therefore,
the presence of shattering would enhance our conclusion
that dust processing after cooling is a key to
understanding the dust abundance and the steepness of
extinction curves in elliptical galaxies.
However, in a high pressure environment in the
central part of elliptical galaxies, such a diffuse
warm medium is not realized as an equilibrium state
\citep{wolfire95}. Therefore, a dynamical model is
necessary to treat this intermediate temperature range
$\sim 10^4$ K, which is left for the future work.

\subsection{Implication of the internal origin of dust}

We have shown that accretion in the cold medium is
a viable mechanism of solving the discrepancy between
the expected dust abundance after sputtering and the
observed one in elliptical galaxies.
Moreover, the steepening of extinction curves
by accretion in the
cooled medium is favoured in terms of the
steepness of observed extinction curves.
Although we do not intend to argue that all the
dust in elliptical galaxies are of internal origin,
it is worth noting that more careful consideration
is necessary before resorting to the external origin
of elliptical galaxy dust.

Even if accretion is the most dominant dust source,
the dust supplied by AGB stars works as seeds for accretion.
Therefore, the dust supply by AGB stars
is still important in obtaining the thorough understanding
of the lifecycle of dust in elliptical galaxies.
The present-day mass loss rate of AGB stars
in elliptical galaxies can be constrained by the mid-infrared
emission emitted by AGB dust \citep[e.g.][]{athey02}.
\citet{villaume15} demonstrated that circumstellar dust around
AGB stars can account for the mid-infrared (8--24 $\micron$)
flux in some early-type galaxies.
Spectral synthesis models including mid-infrared emission of AGB dust
such as those presented in
\citet{villaume15} can be used to constrain the dust yield of
a population of AGB stars in a galaxy using observed dust
emission. This is a crucial component for understanding whether
an internal origin is a plausible explanation for the observed
dust in early-type galaxies.

Theoretical dust yields of AGB stars are also tested
by observations of nearby galaxies. Because of their proximity,
the Magellanic Clouds provide us with an opportunity of detailed
observational constraint on the dust formation in individual
AGB stars \citep{srinivasan09,boyer11,srinivasan11,riebel12,kemper13}.
According to \citet{zhukovska13} and \citet{schneider14},
the current models of dust production in AGB stars
used or discussed in this paper \citep{ferrarotti06,zhukovska08,ventura14}
are broadly consistent with the observed dust production
rate by AGB stars in the
Large Magellanic Cloud. However,
\citet{schneider14} also mentioned that there is still
a discrepancy between the model and observation
for the Small Magellanic Cloud and that the metallicity
dependence of the AGB dust yield is large.
Since dust yields of solar-metallicity AGB stars have not
been tested
against observational data, observational studies on
the dust formation by AGB stars in elliptical galaxies,
which are, broadly speaking, solar-metallicity objects,
give an important constraint on theoretical dust
production models of AGB stars.
The understanding of AGB dust yield is also crucial
even for high-redshift galaxies, in which AGB stars
may contribute to the dust enrichment
\citep{valiante09,gall11,pipino11}.

\section{Conclusion}\label{sec:conclusion}

We have reconsidered the origin of dust in elliptical galaxies,
focusing on the internal origin; that is, the dust originates from
the production by asymptotic giant branch (AGB) stars
within the galaxy. The evolution of grain size distribution is
consistently solved along with the evolution of grain abundance
by dust supply from AGB stars and dust destruction (sputtering)
in the hot interstellar medium (ISM). We have confirmed that
sputtering is so efficient
that the dust abundance observed in the far infrared cannot
be explained. In addition, we have shown that cooling does not
help to protect the dust from sputtering; rather, gas compression
induced by cooling raises the sputtering rate so that the dust-to-gas
ratio is decreased by cooling. Because of the rapid dust destruction,
dust cooling has little influence on the
thermal history of the cooling hot gas, unless the hot gas
initially has a dust-to-gas ratio
as high as $\sim$0.01 (i.e.\ comparable to the value in the Galactic cold ISM).

In the latter part of this paper, we have considered grain growth by the
accretion of gas-phase metals in the cold clouds formed as a result
of cooling of the hot gas, in order to examine a possibility
that the dust abundance is increased or `recovered' by this process.
We find that, if accretion lasts longer than
$10^7/(n_\mathrm{H}/10^3~\mathrm{cm}^{-3})$ yr, which is
shorter than the heating time-scale of active galactic nucleus
feedback, the dust-to-gas ratio can increase up to $\sim 10^{-3}$.
Therefore, we do not necessarily need to resort
to the external origin for the observed excessive dust abundance in
elliptical galaxies, and we suggest dust growth by accretion
in the cooled gas as a plausible explanation for the dust abundance.
Since sputtering in the hot gas destroys small grains more efficiently
than large grains, extinction curves after sputtering are too flat to
explain the observed curves. The observed extinction curves
are better explained by considering the effect of accretion, which
enhances the abundance
of small ($a\la 0.07~\micron$) grains.

In summary, we conclude (i) that cooling does not protect the
dust grains against sputtering, (ii) that dust cooling does
not change the thermal history of the hot gas except for the
case in which the hot gas initially has a dust-to-gas ratio
as large as $\sim$ 0.01, (iii) that the formation of cold
dense gas as a result of cooling helps
dust mass to increase through dust growth by the accretion
of gas-phase metals, and (iv) that the observed
extinction curves are consistent with this accretion
scenario.

\section*{Acknowledgments}

We thank C. Conroy and F. Kemper for valuable discussions.
We also thank the anonymous referee for useful comments.
HH is supported by the Ministry of Science and Technology
(MoST) grant 102-2119-M-001-006-MY3.
TN is supported in part by the JSPS Grant-in-Aid for
Scientific Research (26400223).

\appendix

\section{Evolution of grain size distribution under a varying
background gas density}\label{app:continuity}

In the main text, we only concentrate on the single
fluid element. If we explicitly treat the spatial
variation of grain size distribution by denoting
it as $n(a,\,\bmath{x},\, t)$, where $\bmath{x}$ is the
spatial coordinate, the continuity equation
corresponding to equation (\ref{eq:continuity}) is written
as (see equation 37 in \citealt{tsai95})
\begin{eqnarray}
\left(\frac{\upartial n}{\upartial t}\right)_{a,\bmath{x}}
+\frac{\upartial}{\upartial a}(\dot{a}n)
+\bmath{\nabla}\cdot (n\bmath{u})=S,
\label{eq:continuity_dust}
\end{eqnarray}
where $\bmath{u}$ is the velocity of the fluid,
and $(\upartial /\upartial t)_{a,\bmath{x}}$
is the partial derivative for time with fixed $a$ and $\bmath{x}$.
The gas density follows the continuity equation
written in Lagrangian (comoving) form as
\begin{eqnarray}
\frac{\mathrm{d}\rho_\mathrm{gas}}{\mathrm{d}t}+
\rho_\mathrm{gas}\bmath{\nabla}\cdot\bmath{u}=0.
\label{eq:continuity_gas}
\end{eqnarray}
Eliminating $\bmath{\nabla}\cdot\bmath{u}$ from
equations (\ref{eq:continuity_dust}) and (\ref{eq:continuity_gas}),
we obtain
\begin{eqnarray}
\left(\frac{\upartial n}{\upartial t}\right)_{a,\bmath{x}}
+(\bmath{u}\cdot\bmath{\nabla})n-n\frac{1}{\rho_\mathrm{gas}}
\frac{\mathrm{d}\rho_\mathrm{gas}}{\mathrm{d}t}
+\frac{\upartial}{\upartial a}(\dot{a}n)=S.
\end{eqnarray}
The first two terms are combined to obtain the
Lagrangian derivative (with $a$ fixed):
\begin{eqnarray}
\left(\frac{\upartial n}{\upartial t}\right)_{a}
-n\frac{1}{\rho_\mathrm{gas}}
\frac{\mathrm{d}\rho_\mathrm{gas}}{\mathrm{d}t}
+\frac{\upartial}{\upartial a}(\dot{a}n)=S.
\end{eqnarray}
 Thus, we obtain
equation (\ref{eq:continuity}).

\section{Equilibrium grain size distribution}\label{app:equilibrium}

As seen in the text, the sputtering time-scale is often much shorter than
the cooling time-scale. Therefore, it is useful to discuss the equilibrium
grain size distribution achieved by the balance between sputtering and
supply from AGB stars (we neglect cooling here). Under this equilibrium condition,
equation (\ref{eq:continuity}) is reduced to
\begin{eqnarray}
\frac{\upartial}{\upartial a}[\dot{a}n(a)]=S(a),
\end{eqnarray}
where we set all the time derivative terms as zero.
Considering that $\lim_{a\to\infty}n(a)=0$, the equilibrium
grain size distribution is analytically written as
\begin{eqnarray}
n(a)=\frac{1}{\dot{a}}\int_\infty^aS(a)\,\mathrm{d}a.
\label{eq:eq_ana}
\end{eqnarray}
If we adopt the MRN grain size distribution
(equation \ref{eq:mrn}), we obtain
\begin{eqnarray}
n(a)=\frac{C\dot{\rho}_\mathrm{dust}}{-2.5\dot{a}}
\left( a^{-2.5}-a_\mathrm{max}^{-2.5}\right) ,
\end{eqnarray}
where $a_\mathrm{max}=0.25~\micron$.
Here, we have used equation (\ref{eq:source}) for $S$,
and we can use equation (\ref{eq:sput}) for $\dot{a}$
(note that $\dot{a}$ is negative). If $a$ is significantly
smaller than $a_\mathrm{max}$, $n(a)\propto a^{-2.5}$.
The grain size distribution is less steep than that of
AGB dust, since smaller grains are more easily destroyed
than larger grains.

\begin{figure}
\includegraphics[width=0.45\textwidth]{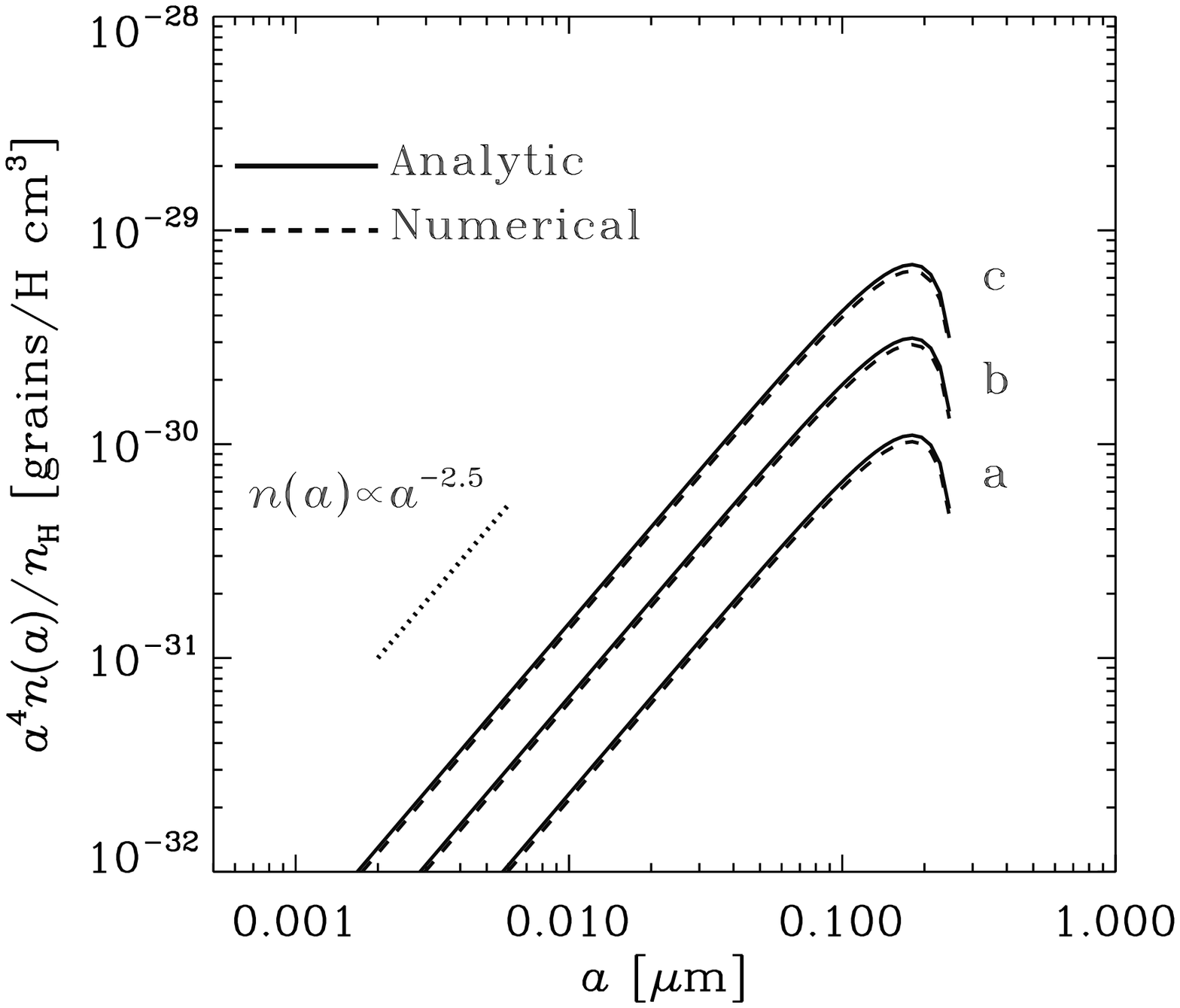}
\caption{Equilibrium grain size distributions for Models
a, b, and c (lower, middle and upper lines, respectively).
We adopt the MRN grain size distribution for AGB dust
and neglect cooling (i.e.\ the gas temperature and density
are constant). The solid and dashed lines show the analytic and
numerical solutions, respectively. For the numerical
solution, we plot the grain size distributions at $t=10^7$ yr,
which is much longer than the sputtering time-scale.
\label{fig:equilibrium}}
\end{figure}

In Fig.\ \ref{fig:equilibrium}, we show the above analytic
solution. For comparison, we also show the numerical
solution of equation (\ref{eq:continuity}) without cooling
at $t=10^7$ yr. Since  the time-scales of
sputtering and dust supply are much shorter than $10^7$
yr, we expect that the grain size distribution has achieved
the equilibrium at $t=10^7$ yr. The analytic and numerical
solutions are almost identical, which confirms that our
numerical scheme correctly solves the evolution of grain
size distribution by dust supply and sputtering.
Moreover, as expected above, the slope of the equilibrium
grain size distribution is described by a power law with
an index of $-2.5$.

\begin{figure}
\includegraphics[width=0.45\textwidth]{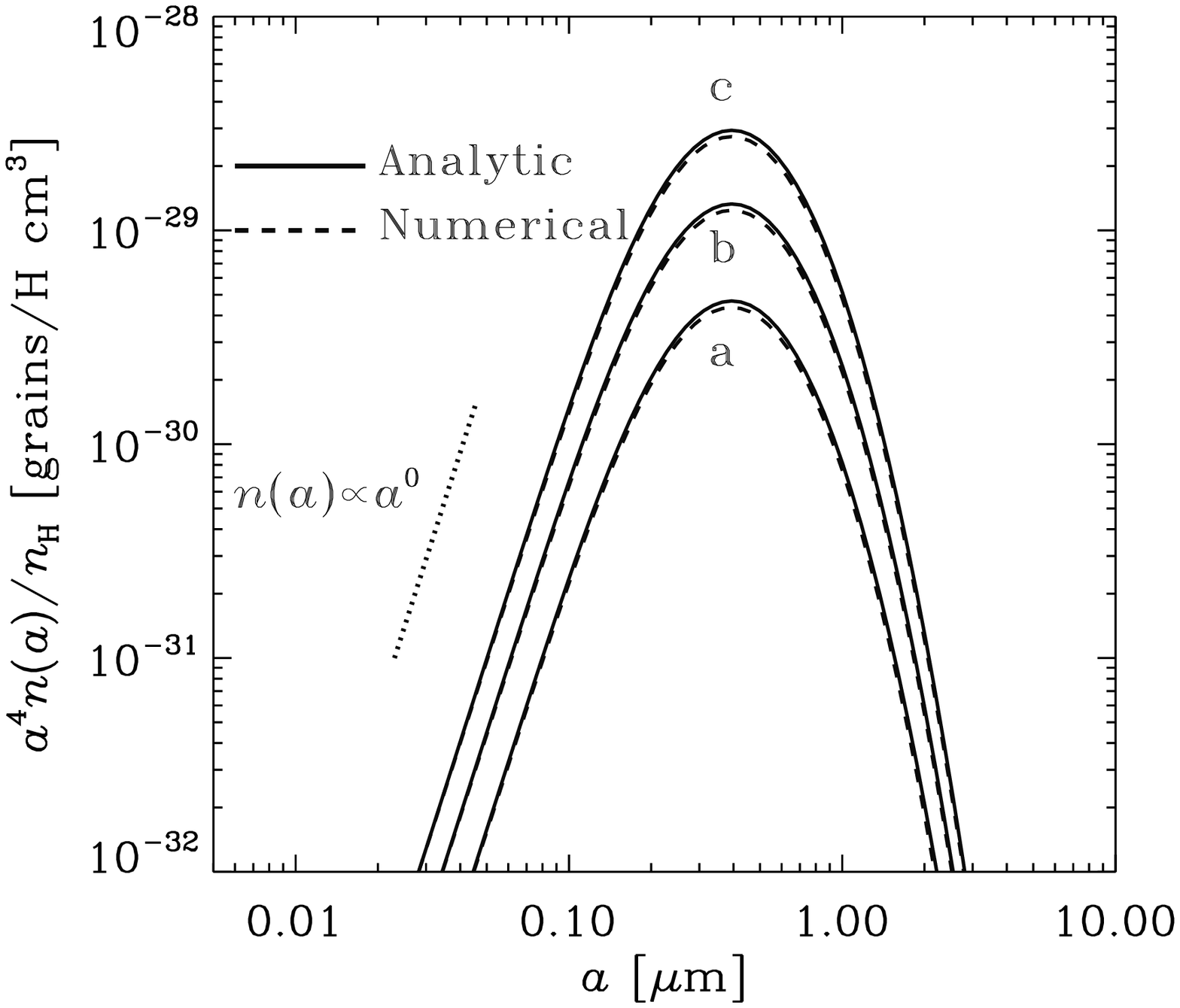}
\caption{Same as Fig.\ \ref{fig:equilibrium} but
for the lognormal grain size distribution for AGB dust.
\label{fig:equil_lognorm}}
\end{figure}

We also show the equilibrium solution for the lognormal
grain size distribution of AGB dust in Fig.\ \ref{fig:equil_lognorm}.
The numerical solution is produced in the same way as above;
that is,
we used the result at $t=10^7$ yr with cooling neglected.
The `analytical' solution is obtained by numerically
integrate equation (\ref{eq:eq_ana}). We confirm that
the numerical results reproduce the analytical results
well. At $a\ll a_0=0.1~\micron$ (the central radius of
the lognormal distribution), $n(a)=\mbox{constant}$
since $S(a)$ drops exponentially in that grain radius
range. We can confirm this in Fig.\ \ref{fig:equil_lognorm}.

\bsp

\label{lastpage}

\end{document}